\documentclass[aps,prb,twocolumn,floatfix]{revtex4}
\usepackage{bm}
\usepackage{tabularx}
\usepackage{ascmac}
\usepackage{amsmath}
\usepackage{float}
\usepackage{wrapfig}
\usepackage{graphicx}
\usepackage[FIGBOTCAP]{subfigure}
\usepackage{color}

\bmdefine{\boldb}{b}
\bmdefine{\bolds}{s}
\bmdefine{\boldS}{S}
\bmdefine{\boldi}{i}
\bmdefine{\boldj}{j}
\bmdefine{\boldl}{l}
\bmdefine{\boldH}{H}
\bmdefine{\boldL}{L}
\bmdefine{\boldJ}{J}
\bmdefine{\boldx}{x}
\bmdefine{\boldX}{X}
\bmdefine{\boldk}{k}
\bmdefine{\boldK}{K}
\bmdefine{\boldp}{p}
\bmdefine{\bolde}{e}
\bmdefine{\boldq}{q}
\bmdefine{\boldQ}{Q}
\bmdefine{\boldD}{D}
\bmdefine{\boldr}{r}
\bmdefine{\boldR}{R}
\bmdefine{\boldm}{m}
\bmdefine{\boldl}{l}
\bmdefine{\boldn}{n}
\bmdefine{\boldv}{v}
\bmdefine{\boldA}{A}
\bmdefine{\boldzero}{0}
\bmdefine{\boldone}{1}

\begin{document}


\title{
Inplane anisotropy of longitudinal thermal conductivities and 
\\weak localization of magnons 
in a disordered spiral magnet
}


\author{Naoya Arakawa}
\email{naoya.arakawa@sci.toho-u.ac.jp} 
\author{Jun-ichiro Ohe}
\affiliation{
Department of Physics, Toho University, 
Funabashi, Chiba, 274-8510, Japan}


\begin{abstract}
We demonstrate 
the inplane anisotropy of longitudinal thermal conductivities and 
the weak localization of magnons  
in a disordered screw-type spiral magnet on a square lattice. 
We consider a disordered spin system, 
described by a spin Hamiltonian for the antiferromagnetic Heisenberg interaction 
and the Dzyaloshinsky-Moriya interaction 
with the mean-field type potential of impurities. 
We derive longitudinal thermal conductivities 
for the disordered screw-type spiral magnet 
in the weak-localization regime 
by using the linear-response theory with the linear-spin-wave approximation 
and performing perturbation calculations. 
We show that 
the inplane longitudinal thermal conductivities are anisotropic 
due to the Dzyaloshinsky-Moriya interaction. 
This anisotropy may be useful for experimentally estimating 
the magnitude of a ratio of the Dzyaloshinsky-Moriya interaction 
to the Heisenberg interaction. 
We also show that 
the main correction term gives 
a logarithmic suppression with the length scale 
due to the critical back scattering. 
This suggests that 
the weak localization of magnons is 
ubiquitous for the disordered two-dimensional magnets 
having global time-reversal symmetry. 
We finally discuss several implications for further research. 

\end{abstract}

\date{\today}
\maketitle


\section{Introduction}

Weak localization of magnons can occur 
in a disordered collinear antiferromagnet~\cite{Lett}. 
A disordered magnet is realized 
by substituting part of magnetic ions in a magnet 
by different ones, which are of the same family in the periodic table~\cite{Lett,Full}. 
This partial substitution modifies the values of exchange interactions, 
and the main effect can be treated as the mean-field type potential~\cite{Lett,Full}. 
Since collinear antiferromagnets have global time-reversal symmetry, 
magnons in disordered two-dimensional collinear antiferromagnets 
will show some characteristic transport properties 
in the weak-localization regime, 
where the effects of disorder can be treated as perturbation. 
(Here
the time-reversal symmetry is defined as the symmetry against 
time-reversal operation for a closed, isolated physical system~\cite{Schiff,Sakurai}; 
our time-reversal symmetry is a global one because 
we have considered not the time-reversal symmetry at a site,
i.e., the local one,  
but the time-reversal symmetry for the system.)
Actually, we demonstrated several properties 
due to the weak localization of magnons~\cite{Lett,Full}. 
For example, 
by treating magnons of disordered Heisenberg antiferromagnets 
in the linear-spin-wave approximation and  
deriving the longitudinal thermal conductivity of magnons  
in the linear-response theory with perturbation calculations, 
we showed that 
the main correction term in the weak-localization regime 
in two dimensions diverges in the thermodynamic limit 
and drastically suppresses the magnon thermal current 
parallel to temperature gradient~\cite{Lett}. 

The results~\cite{Lett} of the disordered collinear antiferromagnet 
provoke two key questions. 
The first one is 
whether the weak localization of magnons occurs 
in other disordered magnets having global time-reversal symmetry; 
the second one is how differences in the magnetic structure 
and exchange interactions 
affect transport properties of disordered magnets. 
These questions will be natural 
because global time-reversal symmetry is vital 
for the weak localization~\cite{Berg,Nagaoka,Lett}, 
and because some magnets, 
such as Ba$_{2}$CuGe$_{2}$O$_{7}$~\cite{CuSpiral1,CuSpiral2,CuSpiral3}, 
have not only the Heisenberg interaction, 
but also the Dzyaloshinsky-Moriya interaction~\cite{DM-D,DM-Moriya}, 
which is absent in the disordered collinear antiferromagnet. 
These questions are also useful for understanding 
the generality of the weak localization of magnons and 
specific properties in each magnet.

To answer these questions, 
we may need to analyze the thermal transport of magnons 
in a disordered screw-type spiral magnet. 
A screw-type spiral magnet~\cite{Yoshimori} has 
the magnetic structure described by, for example, 
$\langle \boldS_{\boldi}\rangle
={}^{t}(S\sin\theta_{\boldi}\ 0\ S\cos\theta_{\boldi})$ 
with  
$\theta_{\boldi}=\boldQ\cdot \boldi$, 
where $S$ is the spin quantum number, 
and $\boldQ$ is the ordering vector. 
Such a screw-type spiral state 
becomes the most stable ground state 
in a spin model for the antiferromagnetic Heisenberg interaction 
and the Dzyaloshinsky-Moriya interaction on a square lattice~\cite{NA-Ir}. 
Then 
the screw-type spiral magnet has global time-reversal symmetry 
because 
its magnetic structure can be regarded as 
a set of antiferromagnetic-like pairs with different, relative angles 
(i.e., a set of the pair for $\langle \boldS_{\boldzero}\rangle$ 
and $-\langle \boldS_{\boldzero}\rangle$, 
the pair for $\langle \boldS_{\boldone}\rangle$ 
and $-\langle \boldS_{\boldone}\rangle$, etc.). 
This property is reasonable 
because the collinear Heisenberg antiferromagnet has global time-reversal symmetry 
and the Dzyaloshinsky-Moriya interaction is symmetric 
about time reversal~\cite{remark}. 
Thus 
a disordered screw-type spiral magnet is suitable 
for comparison with the disordered collinear antiferromagnet. 

However, 
there is no theoretical study about magnon transport  
in the disordered screw-type spiral magnet. 
Such a study is needed to justify the weak localization of magnons 
and understand the effects of the different magnetic structure and exchange interactions. 
Although there is a previous theoretical study~\cite{Evers} about 
the effect of the Dzyaloshinsky-Moriya interaction in a disordered magnet, 
this magnet lacks global time-reversal symmetry.  
It may be desirable to study thermal transport properties of magnons 
in the disordered screw-type spiral magnet. 
This is because the back scattering is critical 
only in the presence of time-reversal symmetry~\cite{Full,Nagaoka}, 
because the critical back scattering is not sufficient to justify 
the weak localization and 
for the justification an analysis of a transport property is necessary. 
Here 
the critical back scattering means 
the divergence of the particle-particle-type four-point vertex function 
in the limit $|\boldQ|=|\boldq+\boldq^{\prime}|\rightarrow 0${\color{red}.} 
Note that in a three-dimensional disordered metal 
the correction term to the longitudinal conductivity in the weak-localization regime 
approaches zero in the thermodynamic limit, 
although the back scattering is critical~\cite{Nagaoka}. 
In the situation where 
the back scattering is critical, 
it is also coherent (for the details, see Appendix A).

In this paper 
we study longitudinal thermal conductivities 
for a disordered two-dimensional spiral magnet in the weak-localization regime. 
The aims of this paper are 
to clarify effects of the Dzyaloshinsky-Moriya interaction, 
which is absent in the disordered antiferromagnet~\cite{Lett,Full}, 
and to justify whether the weak localization of magnons occurs 
in another disordered magnet having global time-reversal symmetry. 
Our spin Hamiltonian includes 
the antiferromagnetic Heisenberg interaction and 
the Dzyaloshinsky-Moriya interaction 
on a square lattice on a $xz$ plane. 
We take account of the main effect of the partial substitution of magnetic ions 
by the mean-field type potential. 
Treating magnons in the linear-spin-wave approximation~\cite{NA-pyro,Gingras,Toth} 
and using the linear-response theory and several approximations 
used for the disordered antiferromagnet~\cite{Lett}, 
we derive the longitudinal thermal conductivities of magnons 
for the disordered screw-type spiral magnet 
in the weak-localization regime. 
We show that 
the inplane longitudinal thermal conductivities are anisotropic 
due to the Dzyaloshinsky-Moriya interaction, 
which results in the difference between magnon propagation 
parallel and perpendicular to the spiral axis.  
We also show that 
the weak localization of magnons occurs 
in the disordered screw-type spiral magnet. 
Then 
we compare the properties of the disordered spiral magnet 
with those of the disordered antiferromagnet 
and discuss the validity of our approximation and the 
implications for further theoretical or experimental studies.

\section{Model}

The Hamiltonian of our model consists of two parts: 
\begin{align}
\hat{H}=\hat{H}_{0}+\hat{H}_{\textrm{imp}},\label{eq:H}
\end{align}
where $\hat{H}_{0}$ is the Hamiltonian without impurities, and 
$\hat{H}_{\textrm{imp}}$ is the Hamiltonian of impurities. 
In the remaining part of this section, 
we first explain the details of $\hat{H}_{0}$, and then 
the details of $\hat{H}_{\textrm{imp}}$. 
For $\hat{H}_{0}$ and $\hat{H}_{\textrm{imp}}$ expressed in terms of magnon operators, 
see Eqs. (\ref{eq:H0-mag-q}) and (\ref{eq:Himp}). 
Throughout this paper we set $\hbar=1$ and $k_{\textrm{B}}=1$.

\subsection{$\hat{H}_{0}$}
\begin{figure}[tb]
\includegraphics[width=66mm]{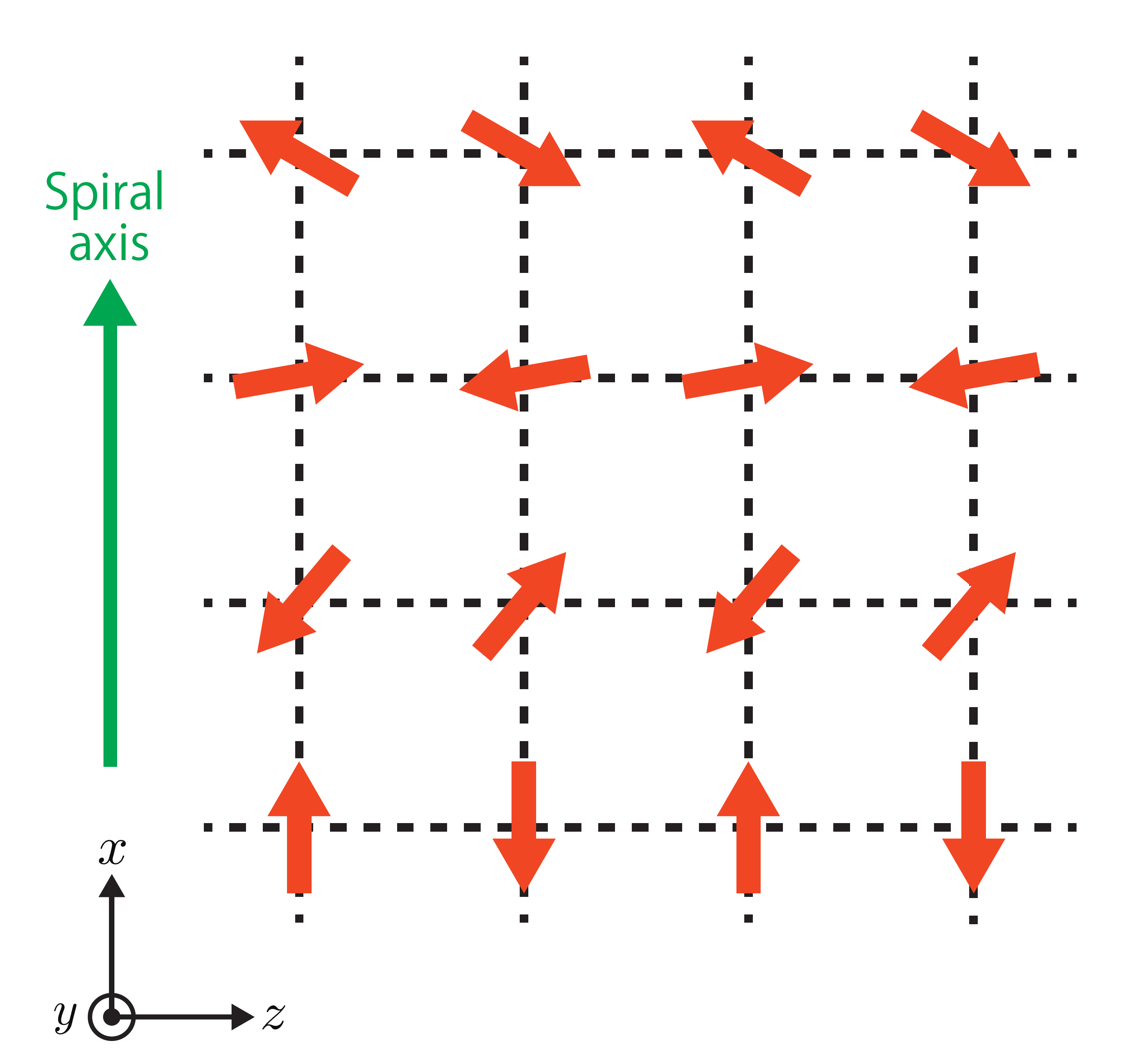}
\caption{
Schematic illustration of the magnetic structure 
of the screw-type spiral magnet. 
The red arrows represent the spins. 
The spin alignment in a $x$ direction is spiral, 
and that in a $z$ direction is antiferromagnetic; 
because of this, 
we call a $x$ axis a spiral axis, which is represented by a green arrow.  
}
\label{fig1}
\end{figure}

As $\hat{H}_{0}$, 
we consider the Heisenberg interaction and Dzyaloshinsky-Moriya interaction 
between nearest-neighbor magnetic ions on a square lattice 
on a $xz$ plane:
\begin{align}
\hat{H}_{0}
=&\sum\limits_{\langle \boldi,\boldj\rangle}J_{\boldi\boldj}
\hat{\boldS}_{\boldi}\cdot \hat{\boldS}_{\boldj}
-\sum\limits_{\langle \boldi,\boldj\rangle}D_{\boldi \boldj}
\Bigl(
\hat{S}_{\boldi}^{z}\hat{S}_{\boldj}^{x}-\hat{S}_{\boldi}^{x}\hat{S}_{\boldj}^{z}
\Bigr)\notag\\
=&\sum\limits_{\boldi,\boldj}\sum\limits_{\alpha,\beta=x,y,z}
M_{\alpha\beta}(\boldi,\boldj)\hat{S}^{\alpha}_{\boldi}\hat{S}^{\beta}_{\boldj}
.\label{eq:H0-spin}
\end{align}
Here $\sum_{\langle \boldi,\boldj\rangle}=\frac{1}{2}\sum_{\boldi,\boldj}$ 
is the summation for nearest-neighbor magnetic ions 
at $\boldi={}^{t}(i_{x}\ i_{z})$ and $\boldj={}^{t}(j_{x}\ j_{z})$ 
on the square lattice; 
$J_{\boldi\boldj}$ is the antiferromagnetic Heisenberg interaction, given by
\begin{align}
J_{\boldi \boldj}=
\begin{cases}
J \ \ (|j_{x}-i_{x}|=1,\ i_{z}=j_{z}),\\
J \ \ (|j_{z}-i_{z}|=1,\ i_{x}=j_{x}),\\
0 \ \ \ (\textrm{otherwise}),
\end{cases}\label{eq:J}
\end{align}
where $J>0$;
$D_{\boldi \boldj}$ is the Dzyaloshinsky-Moriya interaction, given by
\begin{align}
D_{\boldi \boldj}=
\begin{cases}
+D \ \ (j_{x}-i_{x}= +1,\ i_{z}=j_{z}),\\
-D \ \ (j_{x}-i_{x}= -1,\ i_{z}=j_{z}),\\
0 \ \ \ \ \ (\textrm{otherwise}),
\end{cases}\label{eq:DM}
\end{align}
where $D>0$. 
We use Eq. (\ref{eq:H0-spin}) as the Hamiltonian without impurities 
because this is a minimal model for a screw-type spiral magnet. 
As shown in Appendix B, 
the most stable ground state in the mean-field approximation is a screw-type spiral magnet, 
characterized by
\begin{align} 
\langle \boldS_{\boldi}\rangle
=
\left(
\begin{array}{@{\,}c@{\,}}
\langle S^{x}_{\boldi}\rangle\\
\langle S^{y}_{\boldi}\rangle\\
\langle S^{z}_{\boldi}\rangle
\end{array}
\right)
=
\left(
\begin{array}{@{\,}c@{\,}}
S\sin \boldQ\cdot\boldi\\
0\\
S\cos \boldQ\cdot\boldi
\end{array}
\right),
\end{align} 
where $\boldQ={}^{t}(Q_{x}\ Q_{z})$ 
with $Q_{x}=\pi-\cos^{-1}(J/\sqrt{J^{2}+D^{2}})$ and $Q_{z}=\pi$; 
the ground-state energy of this state is always lower than 
that of the antiferromagnetic state for $\boldq={}^{t}(\pi \ \pi)$ 
as long as $D$ is finite. 
The magnetic structure is schematically illustrated in Fig. \ref{fig1}.  

To describe magnon properties of $\hat{H}_{0}$, 
we express $\hat{H}_{0}$ in terms of magnon operators by using 
the linear-spin-wave approximation. 
For the details of the linear-spin-wave approximation 
for a noncollinear magnet, 
see Refs. \onlinecite{Gingras}, \onlinecite{Toth} and \onlinecite{NA-pyro}. 
First, 
we introduce a rotation matrix, defined as follows: 
\begin{align}
\langle \hat{\boldS}_{\boldi}\rangle
=R_{\boldi}\langle \hat{\boldS}^{\prime}_{\boldi}\rangle,
\end{align}
where $\langle \hat{\boldS}^{\prime}_{\boldi}\rangle={}^{t}(0\ 0\ S)$. 
We have introduced this matrix because 
the Holstein-Primakoff transformation for a collinear ferromagnet 
is applicable to $\hat{H}_{0}$ expressed in terms of $\hat{\boldS}^{\prime}_{\boldi}$.
For the screw-type spiral magnet, $R_{\boldi}$ is 
\begin{align}
R_{\boldi}
=
\left(
\begin{array}{@{\,}ccc@{\,}}
\cos(\boldQ\cdot \boldi) & 0 & \sin(\boldQ\cdot\boldi)\\
0 & 1 & 0\\
-\sin(\boldQ\cdot\boldi) & 0 & \cos(\boldQ\cdot \boldi)
\end{array}
\right).
\end{align}
By using this rotation matrix, 
we obtain the following relation between the spin operators: 
\begin{align}
&\hat{S}_{\boldi}^{x}=
\cos(\boldQ\cdot\boldi)\hat{S}_{\boldi}^{\prime x}
+\sin(\boldQ\cdot\boldi)\hat{S}_{\boldi}^{\prime z},\label{eq:Sx'}\\
&\hat{S}_{\boldi}^{y}=\hat{S}_{\boldi}^{\prime y},\label{eq:Sy'}\\
&\hat{S}_{\boldi}^{z}=
-\sin(\boldQ\cdot\boldi)\hat{S}_{\boldi}^{\prime x}
+\cos(\boldQ\cdot\boldi)\hat{S}_{\boldi}^{\prime z}.\label{eq:Sz'}
\end{align}
Second, by using Eqs. (\ref{eq:Sx'}){--}(\ref{eq:Sz'}), 
we express $\hat{H}_{0}$ in terms of $\hat{\boldS}_{\boldi}^{\prime}$. 
As a result, 
we obtain 
\begin{align}
\hat{H}_{0}
=&-\sum\limits_{\langle\boldi,\boldj\rangle}\tilde{J}_{\boldi\boldj}
(\hat{S}_{\boldi}^{\prime x}\hat{S}_{\boldj}^{\prime x}+\hat{S}_{\boldi}^{\prime z}\hat{S}_{\boldj}^{\prime z})
+\sum\limits_{\langle\boldi,\boldj\rangle}J_{\boldi\boldj}
\hat{S}_{\boldi}^{\prime y}\hat{S}_{\boldj}^{\prime y},\label{eq:H0-S'}
\end{align}
where 
\begin{align}
\tilde{J}_{\boldi\boldj}
=
\begin{cases}
\sqrt{J^{2}+D^{2}} \ \ (|j_{x}-i_{x}|=1,\ i_{z}=j_{z}),\\
J \ \ \ \ \ \ \ \ \ \ \ \ \ (|j_{z}-i_{z}|=1,\ i_{x}=j_{x}),\\
0 \ \ \ \ \ \ \ \ \ \ \ \ \ \ (\textrm{otherwise}).
\end{cases}
\end{align}
The details of this derivation are described in Appendix C. 
Third, we express $\hat{H}_{0}$ in terms of magnon operators 
by using the Holstein-Primakoff transformation, 
which connects spin operators and magnon operators as follows:
\begin{align}
&\hat{S}_{\boldi}^{\prime z}=S-\hat{b}^{\dagger}_{\boldi}\hat{b}_{\boldi},\\
&\hat{S}_{\boldi}^{\prime x}=\sqrt{\frac{S}{2}}(\hat{b}_{\boldi}+\hat{b}_{\boldi}^{\dagger}),\\
&\hat{S}_{\boldi}^{\prime y}=-i\sqrt{\frac{S}{2}}(\hat{b}_{\boldi}-\hat{b}_{\boldi}^{\dagger}),
\end{align}
where $\hat{b}^{\dagger}_{\boldi}$ and $\hat{b}_{\boldi}$ 
are creation and annihilation operators of a magnon. 
(Because of this transformation, 
the vectorial nature of spin waves, which are characterized 
as $\Delta \hat{\boldS}_{\boldi}^{\prime}
=\hat{\boldS}_{\boldi}^{\prime}-\langle\hat{\boldS}_{\boldi}^{\prime}\rangle$, 
can be taken into account in the theory using the magnon operators.) 
Since only the quadratic terms of magnon operators 
are considered in the linear-spin-wave approximation, 
the magnon Hamiltonian without impurities for the screw-type spiral magnet 
in the linear-spin-wave approximation is given by
\begin{align}
\hat{H}_{0}=
&S\sum\limits_{\langle \boldi,\boldj\rangle}
\tilde{J}_{\boldi\boldj}(\hat{b}_{\boldi}^{\dagger}\hat{b}_{\boldi}
+\hat{b}_{\boldj}^{\dagger}\hat{b}_{\boldj})
-\frac{S}{2}\sum\limits_{\langle \boldi,\boldj\rangle}
\tilde{J}_{\boldi\boldj}^{(+)}(\hat{b}_{\boldi}\hat{b}_{\boldj}
+\hat{b}_{\boldi}^{\dagger}\hat{b}_{\boldj}^{\dagger})\notag\\
&-\frac{S}{2}\sum\limits_{\langle \boldi,\boldj\rangle}
\tilde{J}_{\boldi\boldj}^{(-)}(\hat{b}_{\boldi}^{\dagger}\hat{b}_{\boldj}
+\hat{b}_{\boldi}\hat{b}_{\boldj}^{\dagger}),\label{eq:H0-mag-site}
\end{align}
where $\tilde{J}_{\boldi\boldj}^{(\pm)}=\tilde{J}_{\boldi\boldj}\pm J_{\boldi\boldj}$. 

Then 
we can obtain the energy dispersion relation of magnon bands 
for our spiral magnet 
by using the Fourier transformations and the Bogoliubov transformation. 
By using the Fourier transformations of the magnon operators 
in Eq. (\ref{eq:H0-mag-site}), e.g., 
$\hat{b}_{\boldi}=\frac{1}{\sqrt{N}}\sum\textstyle_{\boldq}\hat{b}_{\boldq}e^{-i\boldq\cdot\boldi}$, 
we obtain
\begin{align}
\hat{H}_{0}
=
&\sum\limits_{\boldq}A(\boldq)
(\hat{b}_{\boldq}^{\dagger}\hat{b}_{\boldq}+\hat{b}_{-\boldq}\hat{b}_{-\boldq}^{\dagger})
+\sum\limits_{\boldq}B(\boldq)
(\hat{b}_{-\boldq}\hat{b}_{\boldq}+\hat{b}_{\boldq}^{\dagger}\hat{b}_{-\boldq}^{\dagger})\notag\\
=
&\sum\limits_{\boldq}
\left(
\hat{b}^{\dagger}_{\boldq}\ \hat{b}_{-\boldq}
\right)
\left(
\begin{array}{@{\,}cc@{\,}}
A(\boldq) & B(\boldq)\\
B(\boldq) & A(\boldq)
\end{array}
\right)
\left(
\begin{array}{@{\,}c@{\,}}
\hat{b}_{\boldq}\\
\hat{b}_{-\boldq}^{\dagger}
\end{array}
\right).\label{eq:H0-mag-q}
\end{align}
Here
\begin{align}
&A(\boldq)=\frac{S}{2}\tilde{J}(\boldzero)-\frac{S}{4}\tilde{J}^{(-)}(\boldq),\\
&B(\boldq)=-\frac{S}{4}\tilde{J}^{(+)}(\boldq),
\end{align}
where 
$\tilde{J}(\boldzero)=\sum_{j=1}^{z}\tilde{J}_{\boldr_{i}\boldr_{j}}$ and 
$\tilde{J}^{(\pm)}(\boldq)
=\sum_{j=1}^{z}\tilde{J}_{\boldr_{i}\boldr_{j}}^{(\pm)}
e^{i\boldq\cdot(\boldr_{i}-\boldr_{j})}$, 
with $z$, the coordination number. 
Equation (\ref{eq:H0-mag-q}) can be also expressed as follows: 
\begin{align}
\hat{H}_{0}=\sum\limits_{\boldq}\sum\limits_{a,b=1,2}
H_{ab}(\boldq)
\hat{x}_{\boldq a}^{\dagger}\hat{x}_{\boldq b},
\end{align}
where $\hat{x}_{\boldq 1}=\hat{b}_{\boldq}$, $\hat{x}_{\boldq 2}=\hat{b}_{-\boldq}^{\dagger}$, 
$H_{11}(\boldq)=H_{22}(\boldq)=A(\boldq)$, and 
$H_{12}(\boldq)=H_{21}(\boldq)=B(\boldq)$. 
We can diagonalize Eq. (\ref{eq:H0-mag-q}) 
by using the Bogoliubov transformation, 
\begin{align}
\left(
\begin{array}{@{\,}c@{\,}}
\hat{b}_{\boldq}\\
\hat{b}_{-\boldq}^{\dagger}
\end{array}
\right)
=
\left(
\begin{array}{@{\,}cc@{\,}}
\cosh\theta_{\boldq} & -\sinh\theta_{\boldq}\\
-\sinh\theta_{\boldq} & \cosh\theta_{\boldq}
\end{array}
\right)
\left(
\begin{array}{@{\,}c@{\,}}
\hat{\gamma}_{\boldq}\\
\hat{\gamma}_{-\boldq}^{\dagger}
\end{array}
\right),
\end{align}
where the hyperbolic functions are determined by
\begin{align}
\tanh 2\theta_{\boldq}=\frac{B(\boldq)}{A(\boldq)}.\label{eq:tanh}
\end{align}
The diagonalized Hamiltonian is given by
\begin{align}
\hat{H}_{0}
=
&\frac{1}{2}\sum\limits_{\boldq}\epsilon(\boldq)
(\hat{\gamma}_{\boldq}^{\dagger}\hat{\gamma}_{\boldq}
+\hat{\gamma}_{-\boldq}\hat{\gamma}_{-\boldq}^{\dagger})\notag\\
=
&\frac{1}{2}\sum\limits_{\boldq}
\left(
\hat{\gamma}^{\dagger}_{\boldq}\ \hat{\gamma}_{-\boldq}
\right)
\left(
\begin{array}{@{\,}cc@{\,}}
\epsilon(\boldq) & 0\\
0 & \epsilon(\boldq)
\end{array}
\right)
\left(
\begin{array}{@{\,}c@{\,}}
\hat{\gamma}_{\boldq}\\
\hat{\gamma}_{-\boldq}^{\dagger}
\end{array}
\right),\label{eq:H0-mag-band}
\end{align}
where 
\begin{align}
\epsilon(\boldq)=2\sqrt{A(\boldq)^{2}-B(\boldq)^{2}}. 
\end{align}
The most important property of the energy dispersion relation 
is the degeneracy of magnon bands because 
this degeneracy results from global time-reversal symmetry~\cite{remark2}; 
a similar degeneracy exists in a collinear antiferromagnet~\cite{Lett}. 
(This degeneracy is similar to 
the Kramers degeneracy in an electron system with time-reversal symmetry.) 
For other important properties, see Appendix D. 

\subsection{$\hat{H}_{\textrm{imp}}$}

\begin{figure}[tb]
\includegraphics[width=58mm]{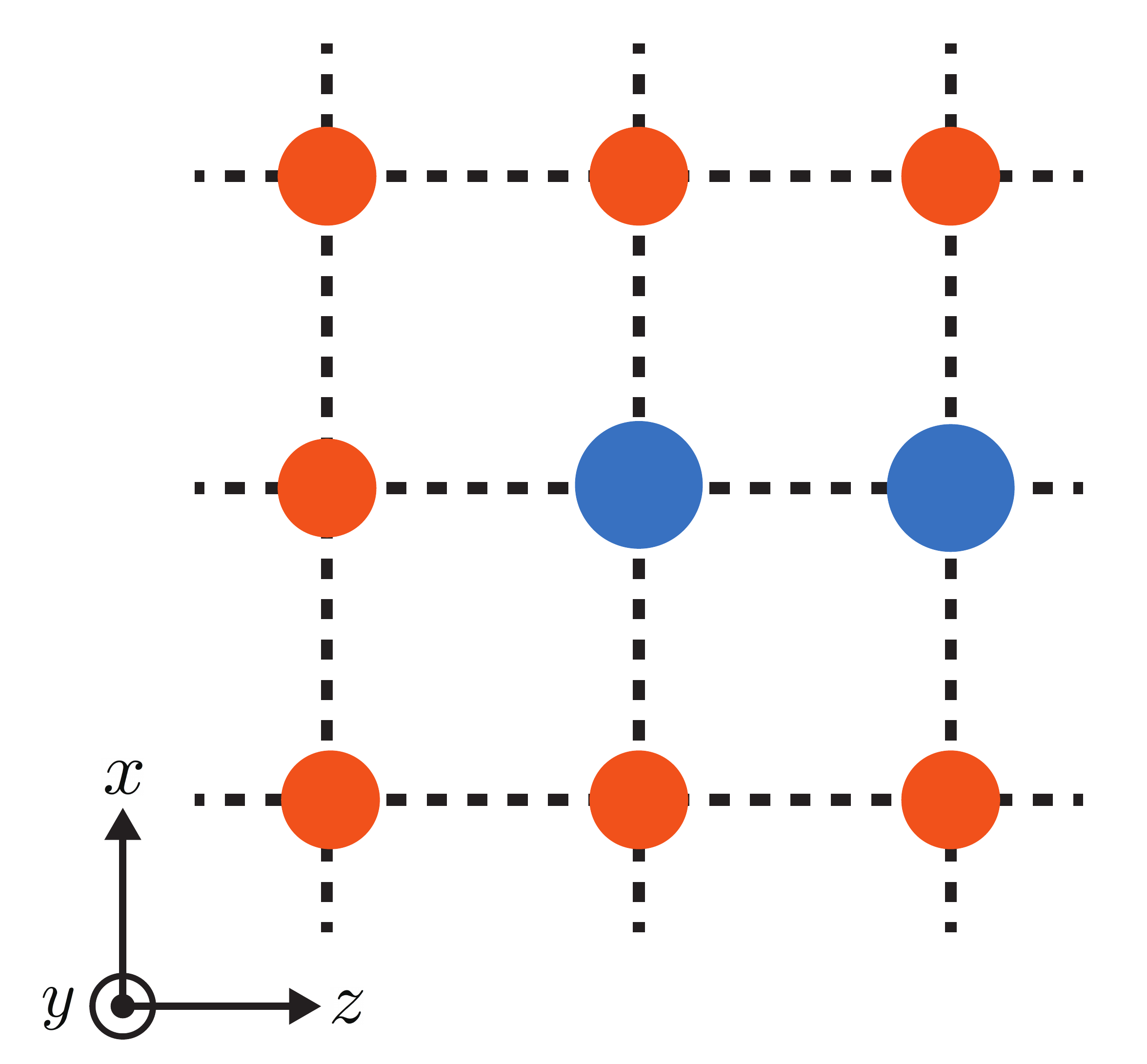}
\caption{
Schematic illustration of our disordered system. 
The orange circles represent original magnetic ions, 
and the blue circles represent impurities, 
which are introduced by substituting part of the original magnetic ions 
by different ones.  
}
\label{fig2}
\end{figure}
We construct $\hat{H}_{\textrm{imp}}$ in a similar way to 
the disordered antiferromagnet~\cite{Lett,Full}. 
We introduce impurities into the screw-type spiral magnet by substituting 
part of the magnetic ions by different ones, 
which belong to the same family in the periodic table; 
our disordered system is schematically illustrated in Fig. \ref{fig2}. 
We have considered such a partial substitution 
because the magnetic ions in the same family 
have the same $S$ and
because its main effect is to modify the values of exchange interactions~\cite{Lett,Full}. 
For our spiral magnet, described by Eq. (\ref{eq:H0-S'}),
such a modification can be described as the following spin Hamiltonian:
\begin{align}
\hat{H}_{\textrm{imp}}
=&-\sum\limits_{\langle\boldi,\boldj\rangle}\Delta\tilde{J}_{\boldi\boldj}
(\hat{S}_{\boldi}^{\prime x}\hat{S}_{\boldj}^{\prime x}+\hat{S}_{\boldi}^{\prime z}\hat{S}_{\boldj}^{\prime z})
+\sum\limits_{\langle\boldi,\boldj\rangle}\Delta J_{\boldi\boldj}
\hat{S}_{\boldi}^{\prime y}\hat{S}_{\boldj}^{\prime y},\label{eq:Himp-S'}
\end{align}
where 
\begin{align}
&\Delta \tilde{J}_{\boldi \boldj}=
\begin{cases}
\tilde{J}^{\prime}_{\boldi\boldj} \ \ \ \ (\boldi\in N_{0},\ \boldj\in N_{\textrm{imp}}),\\
\tilde{J}^{\prime}_{\boldi\boldj} \ \ \ \ (\boldi\in N_{\textrm{imp}},\ \boldj\in N_{0}),\\
\tilde{J}^{\prime\prime}_{\boldi\boldj} \ \ \ \ (\boldi,\boldj \in N_{\textrm{imp}}),\\
0 \ \ \ \ \ \ (\textrm{otherwise}),
\end{cases}\label{eq:Del-tildJ}\\
&\Delta J_{\boldi \boldj}=
\begin{cases}
J^{\prime} \ \ \ \ (\boldi\in N_{0},\ \boldj\in N_{\textrm{imp}}),\\
J^{\prime} \ \ \ \ (\boldi\in N_{\textrm{imp}},\ \boldj\in N_{0}),\\
J^{\prime\prime} \ \ \ \ (\boldi,\boldj \in N_{\textrm{imp}}),\\
0 \ \ \ \ \ \ (\textrm{otherwise}),
\end{cases}\label{eq:DelJ}
\end{align}
with
\begin{align}
&\tilde{J}^{\prime}_{\boldi\boldj}
=
\begin{cases}
\sqrt{(J^{\prime})^{2}+(D^{\prime})^{2}} \ \ (|j_{x}-i_{x}|=1,\ i_{z}=j_{z}),\\
J^{\prime} \ \ \ \ \ \ \ \ \ \ \ \ \ \ \ \ \ \ \ \ (|j_{z}-i_{z}|=1,\ i_{x}=j_{x}),\\
0 \ \ \ \ \ \ \ \ \ \ \ \ \ \ \ \ \ \ \ \ \ (\textrm{otherwise}),
\end{cases}\\ 
&\tilde{J}^{\prime\prime}_{\boldi\boldj}
=
\begin{cases}
\sqrt{(J^{\prime\prime})^{2}+(D^{\prime\prime})^{2}} \ \ (|j_{x}-i_{x}|=1,\ i_{z}=j_{z}),\\
J^{\prime\prime} \ \ \ \ \ \ \ \ \ \ \ \ \ \ \ \ \ \ \ \ (|j_{z}-i_{z}|=1,\ i_{x}=j_{x}),\\
0 \ \ \ \ \ \ \ \ \ \ \ \ \ \ \ \ \ \ \ \ \ \ (\textrm{otherwise}).
\end{cases}
\end{align}
In Eqs. (\ref{eq:Del-tildJ}) and (\ref{eq:DelJ}) 
$N_{0}$ represents the nonsubstituted magnetic ions
(i.e., orange circles in Fig. \ref{fig2})
and $N_{\textrm{imp}}$ represents the impurities 
(i.e., blue circles in Fig. \ref{fig2}). 
We assume that
$J^{\prime}$ and $J^{\prime\prime}$ are much smaller than $J$, 
and that $D^{\prime}$ and $D^{\prime\prime}$ are much smaller than $D$. 
Owing to these assumptions, 
the effects of $\hat{H}_{\textrm{imp}}$ can be treated as perturbation; 
under these assumptions the effects of $J^{\prime}$, $J^{\prime\prime}$, 
$D^{\prime}$, and $D^{\prime\prime}$ on the spin-spiral angle, 
$\boldQ\cdot \boldi$ of $\langle \boldS_{\boldi}\rangle$, are negligible. 
In addition, 
since the dominant terms of Eq. (\ref{eq:Himp-S'}) come from the mean-field terms 
and magnetic ions in the same family in the periodic table have the same $S$, 
$\hat{H}_{\textrm{imp}}$ can be approximated as follows: 
\begin{align}
\hat{H}_{\textrm{imp}}
=-2\sum\limits_{\boldj\in N}V\hat{S}_{\boldj}^{\prime z}
-2\sum\limits_{\boldj\in N_{\textrm{imp}}}V^{(\textrm{imp})}\hat{S}_{\boldj}^{\prime z},\label{eq:Himp-pre}
\end{align}
where $\sum_{\boldj\in N}$ is the summation for all sites, and 
$\sum_{\boldj\in N_{\textrm{imp}}}$ is the summation for impurity sites; 
$V=(Sz^{\prime}/2)[\sqrt{(J^{\prime})^{2}+(D^{\prime})^{2}}+J^{\prime}]$ 
and $V^{(\textrm{imp})}=(Sz^{\prime\prime}/2)[\sqrt{(J^{\prime\prime})^{2}+(D^{\prime\prime})^{2}}+J^{\prime\prime}]$, 
where $z^{\prime}$ and $z^{\prime\prime}$ are the coordination numbers 
for $\Delta\tilde{J}_{\boldi\boldj}=\tilde{J}_{\boldi\boldj}^{\prime}$ 
and $\tilde{J}_{\boldi\boldj}^{\prime\prime}$, respectively. 
In the derivation of Eq. (\ref{eq:Himp-pre}) 
we have used $\langle \hat{\boldS}_{\boldi}^{\prime}\rangle ={}^{t}(0\ 0\ S)$. 
In a similar way to $\hat{H}_{0}$, 
we can express $\hat{H}_{\textrm{imp}}$ in terms of magnon operators: 
\begin{align}
\hat{H}_{\textrm{imp}}
=2\sum\limits_{\boldq}V\hat{b}_{\boldq}^{\dagger}\hat{b}_{\boldq}
+2\sum\limits_{\boldq,\boldq^{\prime}}V^{(\textrm{imp})}(\boldq-\boldq^{\prime})
\hat{b}_{\boldq}^{\dagger}\hat{b}_{\boldq^{\prime}},\label{eq:Himp-full}
\end{align}
where 
\begin{align}
V^{(\textrm{imp})}(\boldq-\boldq^{\prime})
=\frac{1}{N}\sum\limits_{\boldj\in N_{\textrm{imp}}}V^{(\textrm{imp})}e^{i(\boldq-\boldq^{\prime})\cdot \boldj}.
\end{align}
In the following analyses 
we neglect the first term of Eq. (\ref{eq:Himp-full}) for simplicity 
because its effect is a small, uniform shift of the diagonal terms of $\hat{H}_{0}$ 
in Eq. (\ref{eq:H0-mag-q}), 
i.e., shifting $A(\boldq)$ into $A(\boldq)+V$. 
We thus use the following as the Hamiltonian of impurities:
\begin{align}
\hat{H}_{\textrm{imp}}
=&2\sum\limits_{\boldq,\boldq^{\prime}}V^{(\textrm{imp})}(\boldq-\boldq^{\prime})
\hat{b}_{\boldq}^{\dagger}\hat{b}_{\boldq^{\prime}}\notag\\
=&\sum\limits_{\boldq,\boldq^{\prime}}
\sum\limits_{a=1,2}
V^{(\textrm{imp})}(\boldq-\boldq^{\prime})
\hat{x}_{\boldq a}^{\dagger}\hat{x}_{\boldq^{\prime} a}.\label{eq:Himp}
\end{align}

\section{Linear-response theory}

To analyze magnon transport of the disordered spiral magnet, 
we consider longitudinal thermal conductivities
under local equilibrium with local energy conservation. 
A longitudinal thermal conductivity, $\kappa_{\alpha\alpha}$, 
is defined as 
$j_{E}^{\alpha}=\kappa_{\alpha\alpha}(-\partial_{\alpha}T)$, 
where $(-\partial_{\alpha}T)$ is the temperature gradient along the $\alpha$ axis, 
and $j_{E}^{\alpha}$ is the density of the energy current parallel to the temperature gradient. 
This conductivity is suitable for analyses of the weak localization of magnons 
in the presence of global time-reversal symmetry 
because this can be finite even with global time-reversal symmetry. 
(Note that 
other conductivities, such as the thermal Hall conductivity, 
are not suitable because those can be zero at finite temperatures even without impurities.) 
Because of the local equilibrium, the local temperature can be defined. 
Then, 
because of the local energy conservation, 
the energy current operator can be derived from the following equation~\cite{Mahan-text}:
\begin{align}
\hat{\boldJ}_{E}
=i\sum\limits_{\boldm,\boldn}\boldr_{\boldn}[\hat{h}_{\boldm},\hat{h}_{\boldn}],\label{eq:JE-def}
\end{align}
where $\hat{h}_{\boldj}$ is given by $\hat{H}=\sum_{\boldj}\hat{h}_{\boldj}$. 
By calculating the right-hand side of Eq. (\ref{eq:JE-def}) 
for $\hat{H}=\hat{H}_{0}$, 
we obtain 
the energy current operator of a magnon for our disordered spiral magnet,
\begin{align}
\hat{\boldJ}_{E}
=\sum\limits_{\boldq}
\sum\limits_{a,b=1,2}
\hat{x}^{\dagger}_{\boldq a}\bolde_{ab}(\boldq)\hat{x}_{\boldq b},\label{eq:JE}
\end{align}
where 
\begin{align}
&\bolde_{11}(\boldq)=-\bolde_{22}(\boldq)
=2\frac{\partial B(\boldq)}{\partial \boldq}B(\boldq)
-2\frac{\partial A(\boldq)}{\partial \boldq}A(\boldq),\label{eq:e11}\\
&\bolde_{12}(\boldq)=\bolde_{21}(\boldq)
=-2\frac{\partial B(\boldq)}{\partial \boldq}A(\boldq)
-2\frac{\partial A(\boldq)}{\partial \boldq}B(\boldq).\label{eq:e12}
\end{align}
The details of this derivation are described in Appendix E. 
For the energy current operator 
we have neglected the terms due to the combination of $\hat{H}_{0}$ and $\hat{H}_{\textrm{imp}}$ 
because 
in the weak-localization regime 
these terms will be negligible compared with the terms of Eq. (\ref{eq:JE}).  

By using the linear-response theory for $\kappa_{\alpha\alpha}$, 
we can express $\kappa_{\alpha\alpha}$ as follows:
\begin{align}
\kappa_{\alpha\alpha}=
\frac{1}{T}
\lim\limits_{\omega\rightarrow 0}
\dfrac{K_{\alpha\alpha}^{(\textrm{R})}(\omega)
-K_{\alpha\alpha}^{(\textrm{R})}(0)}{i\omega},
\end{align}
where $K_{\alpha\alpha}^{(\textrm{R})}(\omega)=K_{\alpha\alpha}(i\Omega_{n}\rightarrow \omega+i0+)$,
with $\Omega_{n}=2\pi T n$ $(n=0, \pm 1, \cdots)$ and 
\begin{align}
K_{\alpha\alpha}(i\Omega_{n})
=\frac{1}{N}
\int^{T^{-1}}_{0}d\tau e^{i\Omega_{n}\tau}
\langle \textrm{T}_{\tau}  
\hat{J}_{\textrm{E}}^{\alpha}(\tau)
\hat{J}_{\textrm{E}}^{\alpha}\rangle.\label{eq:K-Matsu}
\end{align}
Substituting Eq. (\ref{eq:JE}) into Eq. (\ref{eq:K-Matsu}) 
and using a technique of the quantum field theory~\cite{AGD,Eliashberg,NA-Ch}, 
we can express $K_{\alpha\alpha}(i\Omega_{n})$ in terms of magnon Green's functions: 
\begin{align}
&K_{\alpha\alpha}(i\Omega_{n})
=\frac{1}{N}\sum\limits_{\boldq,\boldq^{\prime}}\sum\limits_{a,b,c,d}
e_{ab}^{\alpha}(\boldq)e_{cd}^{\alpha}(\boldq^{\prime})\notag\\
&\times T\sum\limits_{m}
\langle D_{da}(\boldq^{\prime},\boldq,i\Omega_{m})
D_{bc}(\boldq,\boldq^{\prime},i\Omega_{m}+i\Omega_{n})
\rangle,\label{eq:K-Matsu-second}
\end{align}
where $D_{ab}(\boldq,\boldq^{\prime},i\Omega_{m})$ is the Green's function of a magnon 
in the Matsubara-frequency representation before taking the impurity averaging. 
Then, 
by calculating the summation over the Matsubara frequency in Eq. (\ref{eq:K-Matsu-second}) 
and carrying out the analytic continuation (i.e., $i\Omega_{n}\rightarrow \omega+i0+$), 
$\kappa_{\alpha\alpha}$ can be expressed as follows:
\begin{align}
\kappa_{\alpha\alpha}=&
\frac{1}{TN}\sum\limits_{\boldq,\boldq^{\prime}}
\sum\limits_{a,b,c,d}e_{ab}^{\alpha}(\boldq)e_{cd}^{\alpha}(\boldq^{\prime})
P\int_{-\infty}^{\infty}\frac{d\epsilon}{2\pi}
\Bigl[-\frac{\partial n(\epsilon)}{\partial \epsilon}\Bigr]
\notag\\
&\times \Bigl[
\langle D^{(\textrm{A})}_{da}(\boldq^{\prime},\boldq,\epsilon) 
D^{(\textrm{R})}_{bc}(\boldq,\boldq^{\prime},\epsilon)\rangle\notag\\
&\ \ \ \ 
-\frac{1}{2}\langle D^{(\textrm{R})}_{da}(\boldq^{\prime},\boldq,\epsilon) 
D^{(\textrm{R})}_{bc}(\boldq,\boldq^{\prime},\epsilon)\rangle\notag\\
&\ \ \ \ 
-\frac{1}{2}\langle D^{(\textrm{A})}_{da}(\boldq^{\prime},\boldq,\epsilon) 
D^{(\textrm{A})}_{bc}(\boldq,\boldq^{\prime},\epsilon)\rangle
\Bigr].\label{eq:kappa-exact}
\end{align}
Here $n(\epsilon)=(e^{\epsilon/T}-1)$ is the Bose distribution function;  
$D^{(\textrm{R})}_{ab}(\boldq,\boldq^{\prime},\epsilon)$ and 
$D^{(\textrm{A})}_{ab}(\boldq,\boldq^{\prime},\epsilon)$ 
are the retarded and advanced Green's functions 
in the real-frequency representation before taking the impurity averaging. 
Equation (\ref{eq:kappa-exact}) provides a starting point 
for formulating an approximate theory in the weak-localization regime. 

\section{Weak-localization theory}

We formulate the weak-localization theory for magnons 
of our disordered spiral magnet. 
The weak-localization theory is an approximate theory 
in the weak-localization regime 
because this takes account of the main effect of impurities~\cite{Berg,Nagaoka,Lett} 
in the weak-localization regime. 
Since in Eq. (\ref{eq:kappa-exact}) the main contribution in the weak-localization regime 
comes from the term including 
$\langle D^{(\textrm{A})}_{da}(\boldq^{\prime},\boldq,\epsilon) 
D^{(\textrm{R})}_{bc}(\boldq,\boldq^{\prime},\epsilon)\rangle$~\cite{Berg,Nagaoka,Lett}, 
$\kappa_{\alpha\alpha}$ can be approximated as follows:
\begin{align}
\kappa_{\alpha\alpha}=&
\frac{1}{TN}\sum\limits_{\boldq,\boldq^{\prime}}
\sum\limits_{a,b,c,d}e_{ab}^{\alpha}(\boldq)e_{cd}^{\alpha}(\boldq^{\prime})
P\int_{-\infty}^{\infty}\frac{d\epsilon}{2\pi}
\Bigl[-\frac{\partial n(\epsilon)}{\partial \epsilon}\Bigr]
\notag\\
&\times \langle D^{(\textrm{A})}_{da}(\boldq^{\prime},\boldq,\epsilon) 
D^{(\textrm{R})}_{bc}(\boldq,\boldq^{\prime},\epsilon)\rangle.\label{eq:kappa-WL-pre}
\end{align}
Then, 
by carrying out the perturbation expansions for the magnon Green's functions 
in Eq. (\ref{eq:kappa-WL-pre}), taking the impurity averaging, 
and considering only the dominant terms, 
$\kappa_{\alpha\alpha}$ can be expressed as follows: 
\begin{align}
\kappa_{\alpha\alpha}
=\kappa_{\alpha\alpha}^{(\textrm{Born})}+\Delta\kappa_{\alpha\alpha},\label{eq:kappa-WL}
\end{align}
where $\kappa_{\alpha\alpha}^{(\textrm{Born})}$ is the term in the Born approximation, 
\begin{align}
\kappa_{\alpha\alpha}^{(\textrm{Born})}
=&
\frac{1}{TN}\sum\limits_{\boldq}
\sum\limits_{a,b,c,d}e_{ab}^{\alpha}(\boldq)e_{cd}^{\alpha}(\boldq)
P\int_{-\infty}^{\infty}\frac{d\epsilon}{2\pi}
\Bigl[-\frac{\partial n(\epsilon)}{\partial \epsilon}\Bigr]
\notag\\
&\times \bar{D}^{(\textrm{A})}_{da}(\boldq,\epsilon) 
\bar{D}^{(\textrm{R})}_{bc}(\boldq,\epsilon),\label{eq:kappa^Born-WL}
\end{align}
and 
$\Delta\kappa_{\alpha\alpha}$ is the main correction term 
in the weak-localization regime, 
\begin{align}
\Delta\kappa_{\alpha\alpha}
=&
\frac{1}{TN}\sum\limits_{\boldq,\boldq^{\prime}}
\sum\limits_{a,b,c,d}
e_{ab}^{\alpha}(\boldq)e_{cd}^{\alpha}(\boldq^{\prime})
P\int_{-\infty}^{\infty}\frac{d\epsilon}{2\pi}
\Bigl[-\frac{\partial n(\epsilon)}{\partial \epsilon}\Bigr]
\notag\\
&\times 
\sum\limits_{a^{\prime},b^{\prime},c^{\prime},d^{\prime}}
\bar{D}^{(\textrm{A})}_{dd^{\prime}}(\boldq^{\prime},\epsilon) 
\bar{D}^{(\textrm{A})}_{a^{\prime}a}(\boldq,\epsilon) 
\Gamma_{a^{\prime}b^{\prime}c^{\prime}d^{\prime}}(\boldq+\boldq^{\prime},\epsilon)\notag\\
&\times
\bar{D}^{(\textrm{R})}_{bb^{\prime}}(\boldq,\epsilon)
\bar{D}^{(\textrm{R})}_{c^{\prime}c}(\boldq^{\prime},\epsilon).\label{eq:Delkappa-WL}
\end{align}
We have introduced the following quantities:  
$\bar{D}^{(\textrm{R})}_{ab}(\boldq,\epsilon)$ and 
$\bar{D}^{(\textrm{A})}_{ab}(\boldq,\epsilon)$ 
are the retarded and advanced Green's functions 
after taking the impurity averaging; 
$\Gamma_{abcd}(\boldQ,\epsilon)$ is 
the particle-particle type four-point vertex function. 
These Green's functions are determined from the Dyson equation 
for the self-energy in the Born approximation; 
for example, 
$\bar{D}^{(\textrm{R})}_{ab}(\boldq,\epsilon)$ is given by 
\begin{align}
\bar{D}^{(\textrm{R})}_{ab}(\boldq,\epsilon)
=&D^{0(\textrm{R})}_{ab}(\boldq,\epsilon)\notag\\
&+\sum\limits_{a^{\prime},b^{\prime}}
D^{0(\textrm{R})}_{aa^{\prime}}(\boldq,\epsilon)
\Sigma_{a^{\prime}b^{\prime}}^{(\textrm{R})}(\epsilon)
\bar{D}^{(\textrm{R})}_{b^{\prime}b}(\boldq,\epsilon),\label{eq:D-Dyson}
\end{align}
with 
\begin{align}
&D^{0(\textrm{R})}_{ab}(\boldq,\epsilon)
=\dfrac{U_{a\alpha}(\boldq)U_{b\alpha}(\boldq)}{\epsilon-\epsilon(\boldq)+i\delta}
-\dfrac{U_{a\beta}(\boldq)U_{b\beta}(\boldq)}{\epsilon+\epsilon(\boldq)+i\delta},\label{eq:D^0}\\
&\Sigma_{ab}^{(\textrm{R})}(\epsilon)
=\dfrac{n_{\textrm{imp}}V_{\textrm{imp}}^{2}}{N}
\sum\limits_{\boldq}\bar{D}^{(\textrm{R})}_{ab}(\boldq,\epsilon),\label{eq:Sig^R}
\end{align}
where $\delta=0+$, 
$U_{1\alpha}(\boldq)=U_{2\beta}(\boldq)=\cosh\theta_{\boldq}$, 
$U_{1\beta}(\boldq)=U_{2\alpha}(\boldq)=-\sinh\theta_{\boldq}$, 
and $n_{\textrm{imp}}$ is the impurity concentration. 
In addition, 
$\Gamma_{abcd}(\boldQ,\epsilon)$ is determined from 
the following Bethe-Salpeter equation: 
\begin{align}
\Gamma_{abcd}(\boldQ,\epsilon)
=&\gamma_{\textrm{imp}}^{2}\Pi_{abcd}(\boldQ,\epsilon)\notag\\
&+\sum\limits_{e,f}\gamma_{\textrm{imp}}
\Pi_{aecf}(\boldQ,\epsilon)
\Gamma_{fbed}(\boldQ,\epsilon),\label{eq:BSeq}
\end{align}
where $\gamma_{\textrm{imp}}=\frac{n_{\textrm{imp}}V_{\textrm{imp}}^{2}}{N}$ and 
\begin{align}
\Pi_{abcd}(\boldQ,\epsilon)
=\sum\limits_{\boldq_{1}}
\bar{D}^{(\textrm{R})}_{bc}(\boldq_{1},\epsilon)
\bar{D}^{(\textrm{A})}_{da}(\boldQ-\boldq_{1},\epsilon).\label{eq:Pi}
\end{align}

Then we introduce two simplifications. 
One is to neglect the real part of the self-energy, 
i.e., consider only the imaginary part; 
as a result, 
\begin{align}
&\Sigma_{ab}^{(\textrm{R})}(\epsilon)=-i\gamma_{ab}(\epsilon),\\
&\Sigma_{ab}^{(\textrm{A})}(\epsilon)=i\gamma_{ab}(\epsilon),
\end{align} 
where 
\begin{align}
\gamma_{ab}(\epsilon)=-\gamma_{\textrm{imp}}\sum\limits_{\boldq}
\textrm{Im}\bar{D}_{ab}^{(\textrm{R})}(\boldq,\epsilon).
\end{align}
The other is to approximate $D^{0(\textrm{R})}_{ab}(\boldq,\epsilon)$ and 
$D^{0(\textrm{A})}_{ab}(\boldq,\epsilon)$ as follows:
\begin{align}
&D^{0(\textrm{R})}_{ab}(\boldq,\epsilon)\sim 
\begin{cases}
\dfrac{U_{a\alpha}(\boldq)U_{b\alpha}(\boldq)}{\epsilon-\epsilon_{\boldq}+i\delta}\ \
(\epsilon > 0),\\[6pt]
-\dfrac{U_{a\beta}(\boldq)U_{b\beta}(\boldq)}{\epsilon+\epsilon_{\boldq}+i\delta}\ (\epsilon < 0),
\end{cases}\\
&D^{0(\textrm{A})}_{ab}(\boldq,\epsilon)\sim 
\begin{cases}
\dfrac{U_{a\alpha}(\boldq)U_{b\alpha}(\boldq)}{\epsilon-\epsilon_{\boldq}-i\delta}\ \ 
(\epsilon > 0),\\[6pt]
-\dfrac{U_{a\beta}(\boldq)U_{b\beta}(\boldq)}{\epsilon+\epsilon_{\boldq}-i\delta}\ (\epsilon < 0).
\end{cases}
\end{align}
These simplifications, which are similar to those for the disordered antiferromagnet~\cite{Lett},  
will be appropriate for a rough estimate of the main effect of impurities 
because the imaginary part of the self-energy is vital 
for the weak localization~\cite{Berg,Nagaoka,Lett}, 
and because the main contribution to $D^{0(\textrm{R})}_{ab}(\boldq,\epsilon)$ 
for $\epsilon > 0$ or $\epsilon < 0$ comes from  
the first or second term, respectively, of Eq. (\ref{eq:D^0}). 

By using the above two simplifications, 
we can obtain the approximate expressions of 
$\bar{D}^{(\textrm{R})}_{ab}(\boldq,\epsilon)$, 
$\bar{D}^{(\textrm{A})}_{ab}(\boldq,\epsilon)$, 
$\Pi_{abcd}(\boldQ,\epsilon)$, and $\Gamma_{abcd}(\boldQ,\epsilon)$.  
First, 
by combining the simplifications with the Dyson equation, 
$\bar{D}^{(\textrm{R})}_{ab}(\boldq,\epsilon)$ and $\bar{D}^{(\textrm{A})}_{ab}(\boldq,\epsilon)$ 
can be expressed as follows:
\begin{align}
&\bar{D}^{(\textrm{R})}_{ab}(\boldq,\epsilon)\sim 
\begin{cases}
\dfrac{U_{a\alpha}(\boldq)U_{b\alpha}(\boldq)}{\epsilon-\epsilon(\boldq)+i\tilde{\gamma}(\epsilon)}
\ \ \ \ \ \ (\epsilon > 0),\\[6pt]
-\dfrac{U_{a\beta}(\boldq)U_{b\beta}(\boldq)}{\epsilon+\epsilon(\boldq)+i\tilde{\gamma}(-\epsilon)}
\ (\epsilon < 0),
\end{cases}\label{eq:D^R}\\
&\bar{D}^{(\textrm{A})}_{ab}(\boldq,\epsilon)\sim 
\begin{cases}
\dfrac{U_{a\alpha}(\boldq)U_{b\alpha}(\boldq)}{\epsilon-\epsilon(\boldq)-i\tilde{\gamma}(\epsilon)}
\ \ \ \ \ \ (\epsilon > 0),\\[6pt]
-\dfrac{U_{a\beta}(\boldq)U_{b\beta}(\boldq)}{\epsilon+\epsilon(\boldq)-i\tilde{\gamma}(-\epsilon)}
\ (\epsilon < 0).
\end{cases}\label{eq:D^A}
\end{align}
with
\begin{align}
\tilde{\gamma}(\epsilon)
=&(\cosh^{2}\theta_{\boldq}+\sinh^{2}\theta_{\boldq})^{2}\gamma(\epsilon)\notag\\
=&(\cosh^{2}\theta_{\boldq}+\sinh^{2}\theta_{\boldq})^{2}
\pi n_{\textrm{imp}}V_{\textrm{imp}}^{2}\rho(\epsilon),
\end{align}
where $\boldq$ of $\cosh^{2}\theta_{\boldq}$ and $\sinh^{2}\theta_{\boldq}$ are determined by 
$\epsilon(\boldq)=|\epsilon|$, 
and $\rho(\epsilon)$ is the density of states. 
Second, 
by substituting Eqs. (\ref{eq:D^R}) and (\ref{eq:D^A}) into Eq. (\ref{eq:Pi}) 
and performing the calculations described in Appendix F, 
we can express $\Pi_{abcd}(\boldQ,\epsilon)$ for small $Q=|\boldQ|$ as follows:
\begin{align}
&\Pi_{abcd}(\boldQ,\epsilon)\notag\\
&\sim 
\begin{cases}
\dfrac{u_{b\alpha}u_{c\alpha}u_{d\alpha}u_{a\alpha}}{\gamma_{\textrm{imp}}(c_{0}^{2}+s_{0}^{2})^{2}}
[1-D_{\textrm{S}}(\epsilon)Q^{2}\tilde{\tau}(\epsilon)]
\ \ \ \ \ (\epsilon > 0),\\[8pt]
\dfrac{u_{b\beta}u_{c\beta}u_{d\beta}u_{a\beta}}{\gamma_{\textrm{imp}}(c_{0}^{2}+s_{0}^{2})^{2}}
[1-D_{\textrm{S}}(-\epsilon)Q^{2}\tilde{\tau}(-\epsilon)]
\ (\epsilon < 0),
\end{cases}\label{eq:Pi-approx}
\end{align}
where 
$u_{a\nu}=U_{a\nu}(\boldq_{0})$ ($\nu=\alpha,\beta$), 
$c_{0}^{2}=\cosh^{2}\theta_{\boldq_{0}}$, 
$s_{0}^{2}=\sinh^{2}\theta_{\boldq_{0}}$, 
$D_{\textrm{S}}(\epsilon)=\frac{1}{8}
[\frac{\partial \epsilon(\boldq_{0})}{\partial \boldq_{0}}]^{2}\tilde{\tau}(\epsilon)
=\frac{1}{8}
\boldv_{\boldq_{0}}^{2}\tilde{\tau}(\epsilon)$, 
and $\tilde{\tau}(\epsilon)=(c_{0}^{2}+s_{0}^{2})^{-2}\gamma(\epsilon)^{-1}$. 
In the derivation of Eq. (\ref{eq:Pi-approx}) 
we have approximated momentum-dependent quantities, 
$U_{a\nu}(\boldq_{1})$ and 
$[\frac{\partial \epsilon(\boldq_{1})}{\partial \boldq_{1}}]$, 
as the typical values at a certain, small momentum $\boldq_{0}$ 
for a rough estimate 
because the dominant contributions come from the contributions for small $q_{1}=|\boldq_{1}|$. 
Note, first, 
that 
the group velocity of the magnon for $\boldq=\boldzero$ is zero in our spiral magnet;  
second, that 
since the small-momentum contributions are dominant in the summation, 
the sum of a function $F(\boldq)$ might be approximated by 
$\sum_{\boldq}F(\boldq)\sim \sum_{0\leq |\boldq|\leq q_{\textrm{c}}}
F(\boldq)\sim F(\boldq_{0})\sum_{0\leq |\boldq|\leq q_{\textrm{c}}}$, 
where $q_{\textrm{c}}$ is a cut-off value. 
(In a rough sense 
this approximation is similar to a replacement of a momentum-dependent quantity 
in an electron system by the quantity at the Fermi momentum.)
We have shown the approximate expression of $\Pi_{abcd}(\boldQ,\epsilon)$ only for small $Q$ 
because the contributions for small $Q$ lead to 
the main contribution to $\Delta\kappa_{\alpha\alpha}$ 
through the diverging contribution of $\Gamma_{abcd}(\boldQ,\epsilon)$ 
for $\boldQ=\boldq+\boldq^{\prime}$. 
Third, 
by combining Eq. (\ref{eq:Pi-approx}) with Eq. (\ref{eq:BSeq}) 
and solving the Bethe-Salpeter equation in the way described in Appendix G, 
we obtain the approximate expression of $\Gamma_{abcd}(\boldQ,\epsilon)$ for small $Q$: 
\begin{align}
\Gamma_{abcd}(\boldQ,\epsilon)\sim 
\begin{cases}
\dfrac{u_{b\alpha}u_{c\alpha}u_{d\alpha}u_{a\alpha}\gamma_{\textrm{imp}}}
{D_{\textrm{S}}(\epsilon)Q^{2}\tilde{\tau}(\epsilon)}
\ \ (\epsilon > 0),\\[8pt]
\dfrac{u_{b\beta}u_{c\beta}u_{d\beta}u_{a\beta}\gamma_{\textrm{imp}}}
{D_{\textrm{S}}(-\epsilon)Q^{2}\tilde{\tau}(-\epsilon)}
\ \ \ (\epsilon < 0).
\end{cases}\label{eq:Gamma-approx}
\end{align}
Since $\Gamma_{abcd}(\boldQ,\epsilon)$ diverges in the limit $Q\rightarrow 0$, 
the particle-particle type multiple scattering, 
described by $\Gamma_{abcd}(\boldQ,\epsilon)$, 
for $\boldQ=\boldq+\boldq^{\prime}=\boldzero$ provides 
the diverging contribution to $\Delta\kappa_{\alpha\alpha}$. 

We can also obtain the approximate expressions of $\kappa_{\alpha\alpha}^{(\textrm{Born})}$ 
and $\Delta\kappa_{\alpha\alpha}$. 
First, 
by combining Eqs. (\ref{eq:kappa^Born-WL}), (\ref{eq:e11}), (\ref{eq:e12}), 
(\ref{eq:D^R}), and (\ref{eq:D^A}), 
we obtain
\begin{align}
\kappa_{\alpha\alpha}^{(\textrm{Born})}=\frac{1}{TN}
\sum\limits_{\boldq}\tilde{e}^{\alpha}(\boldq)^{2}
\Bigl\{-\frac{\partial n[\epsilon(\boldq)]}{\partial \epsilon(\boldq)}\Bigr\}
\tilde{\tau}[\epsilon(\boldq)],
\label{eq:kappa^Born-final}
\end{align}
where 
\begin{align}
\tilde{e}^{\alpha}(\boldq)^{2}=
e_{11}^{\alpha}(\boldq)^{2}+e_{12}^{\alpha}(\boldq)^{2}\sinh^{2}2\theta_{\boldq}.
\end{align}
In deriving Eq. (\ref{eq:kappa^Born-final}) 
we have approximated $[-\frac{\partial n(\epsilon)}{\partial \epsilon}]$ 
and $\tilde{\tau}(\epsilon)$ as 
$\{-\frac{\partial n[\epsilon(\boldq)]}{\partial \epsilon(\boldq)}\}$ 
and $\tilde{\tau}[\epsilon(\boldq)]$ 
because the contributions near $\epsilon=\epsilon(\boldq)$ or $\epsilon=-\epsilon(\boldq)$ 
are dominant for $\epsilon > 0$ or $\epsilon < 0$, respectively. 
Then we can obtain 
the approximate expression of $\Delta\kappa_{\alpha\alpha}$ in the following way. 
To estimate the main effect of 
the diverging contribution of $\Gamma_{abcd}(\boldq+\boldq^{\prime},\epsilon)$, 
we set $\boldq^{\prime}=-\boldq$ in Eq. (\ref{eq:Delkappa-WL}) except for 
$\Gamma_{abcd}(\boldq+\boldq^{\prime},\epsilon)$ 
and introduce the cutoff values for the upper and lower values of 
the summation over $\boldq^{\prime}$; 
the lower cutoff value is $|\boldQ|=|\boldq+\boldq^{\prime}|=L^{-1}$, 
which approaches zero in the thermodynamic limit, 
and the upper cutoff value is $|\boldQ|=|\boldq+\boldq^{\prime}|=L_{\textrm{m}}^{-1}$ 
with the mean-free path $L_{\textrm{m}}$. 
As a result, 
$\Delta\kappa_{\alpha\alpha}$ is given by
\begin{align}
&\Delta\kappa_{\alpha\alpha}
=-\frac{1}{TN}\sum\limits_{\boldq}\sum\limits_{a,b,c,d}
e_{ab}^{\alpha}(\boldq)e_{cd}^{\alpha}(\boldq)P\int_{-\infty}^{\infty}\frac{d\epsilon}{2\pi}
\Bigl[-\frac{\partial n(\epsilon)}{\partial \epsilon}\Bigr]\notag\\
&\times 
\sum\limits_{a^{\prime},b^{\prime},c^{\prime},d^{\prime}}
\bar{D}_{dd^{\prime}}^{(\textrm{A})}(\boldq,\epsilon)
\bar{D}_{a^{\prime}a}^{(\textrm{A})}(\boldq,\epsilon)
\bar{D}_{bb^{\prime}}^{(\textrm{R})}(\boldq,\epsilon)
\bar{D}_{c^{\prime}c}^{(\textrm{R})}(\boldq,\epsilon)\notag\\
&\times \sum\limits_{\boldQ}{}^{\prime}\Gamma_{a^{\prime}b^{\prime}c^{\prime}d^{\prime}}(\boldQ,\epsilon)
\label{eq:Delkapp-pre},
\end{align}
where $\frac{1}{N}\sum_{\boldQ}^{\prime}=\int_{L^{-1}}^{L_{\textrm{m}}^{-1}}\frac{dq}{(2\pi)^{2}}2\pi q$. 
Combining Eqs. (\ref{eq:Delkapp-pre}), (\ref{eq:e11}), (\ref{eq:e12}), 
(\ref{eq:D^R}), (\ref{eq:D^A}), and (\ref{eq:Gamma-approx}), 
we obtain
\begin{align}
\Delta\kappa_{\alpha\alpha}
&\sim -\frac{1}{TN}
\sum\limits_{\boldq}
\tilde{e}^{\alpha}(\boldq)^{2}
\Bigl\{-\frac{\partial n[\epsilon(\boldq)]}{\partial \epsilon(\boldq)}\Bigr\}
\tilde{\tau}[\epsilon(\boldq)]\notag\\
&\times \frac{n_{\textrm{imp}}V_{\textrm{imp}}^{2}}
{8\pi D_{\textrm{S}}[\epsilon(\boldq_{0})]\gamma[\epsilon(\boldq_{0})]}
\frac{(c_{0}-s_{0})^{4}}{(c_{0}^{2}+s_{0}^{2})^{4}}\ln\Bigl(\frac{L}{L_{\textrm{m}}}\Bigr)\notag\\
&=
-\kappa_{\alpha\alpha}^{(\textrm{Born})}
\frac{n_{\textrm{imp}}V_{\textrm{imp}}^{2}}{\pi \tilde{\boldv}_{0}^{2}}
\ln\Bigl(\frac{L}{L_{\textrm{m}}}\Bigr)
\label{eq:Delkappa-final},
\end{align}
where 
\begin{align}
\tilde{\boldv}_{0}^{2}=\boldv_{0}^{2}
\frac{(c_{0}^{2}+s_{0}^{2})^{2}}{(c_{0}-s_{0})^{4}}.
\end{align}
In this derivation we have used the approximations used to derive
Eqs. (\ref{eq:Pi-approx}) and (\ref{eq:kappa^Born-final}). 
Equation (\ref{eq:Delkappa-final}) shows that 
$\Delta\kappa_{\alpha\alpha}$ leads to a negative logarithmic divergence 
in the thermodynamic limit. 
This is the same behavior as 
the weak localization of magnons in the disordered antiferromagnet~\cite{Lett}; 
the differences between the disordered spiral magnet and the disordered antiferromagnet 
are the different expressions of $\kappa_{\alpha\alpha}^{(\textrm{Born})}$ 
and $\tilde{\boldv}_{0}^{2}$. 
We thus conclude that 
the weak localization of magnons occurs in the disordered screw-type spiral magnet 
in two dimensions. 

\begin{figure*}[tb]
\includegraphics[width=160mm]{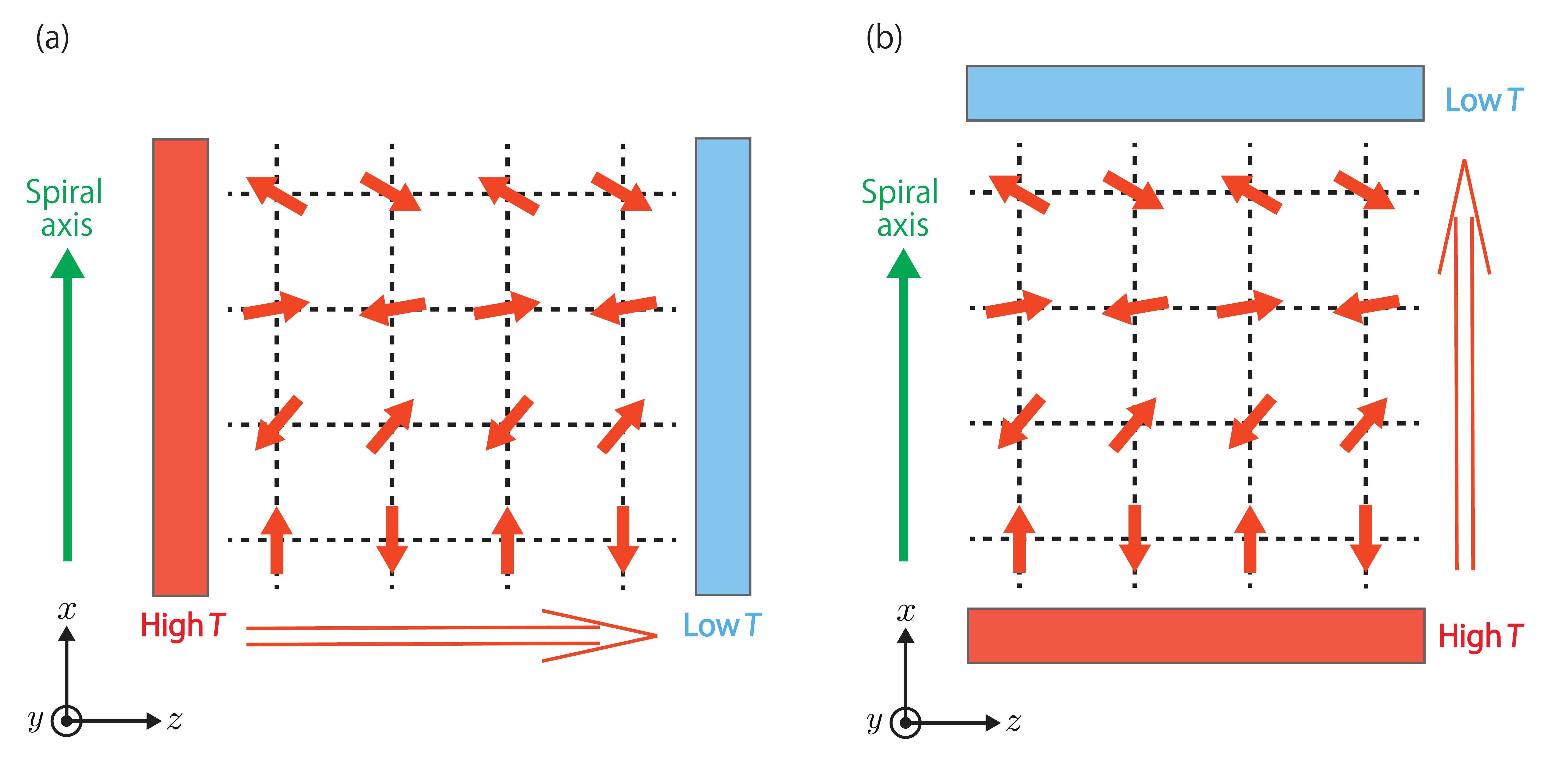}
\caption{
Schematic illustrations of the situations 
for (a) $\kappa_{zz}$ and (b) $\kappa_{xx}$. 
The red rectangles and blue rectangles represent the temperature gradient; 
the green arrows represent the spiral axis; 
the short and long arrows represent the spins and the magnon thermal currents, 
respectively. 
}
\label{fig3}
\end{figure*}

We now turn to the inplane anisotropy of longitudinal thermal conductivities. 
For our screw-type spiral magnet on a $xz$ plane, 
the spiral axis is parallel to a $x$ axis and perpendicular to a $z$ axis, 
as shown in Fig. \ref{fig1}; 
this is because $\langle \boldS_{\boldi}\rangle=
{}^{t}(S\sin \boldQ\cdot\boldi \ 0\ S\cos \boldQ\cdot\boldi)$ 
with $Q_{x}=\pi-\cos^{-1}(J/\sqrt{J^{2}+D^{2}})$ and $Q_{z}=\pi$. 
Since this spiral alignment in a $x$ direction results from 
the combination of the antiferromagnetic Heisenberg interaction 
and the Dzyaloshinsky-Moriya interaction, 
we expect that 
$\kappa_{xx}$ and $\kappa_{zz}$ are different 
and this difference is connected to a ratio of the Dzyaloshinsky-Moriya interaction 
to the Heisenberg interaction. 
To justify this expectation, 
we estimate $\kappa_{xx}/\kappa_{zz}$; 
the situations for $\kappa_{zz}$ and $\kappa_{xx}$ are 
schematically illustrated in Fig. \ref{fig3}. 
From Eqs. (\ref{eq:kappa-WL}), (\ref{eq:kappa^Born-final}), and 
(\ref{eq:Delkappa-final}) 
we have
\begin{align}
\frac{\kappa_{xx}}{\kappa_{zz}}
=&\frac{\kappa_{xx}^{(\textrm{Born})}}{\kappa_{zz}^{(\textrm{Born})}}
=\frac{\sum\limits_{\boldq}\tilde{e}^{x}(\boldq)^{2}
\Bigl\{-\frac{\partial n[\epsilon(\boldq)]}{\partial \epsilon(\boldq)}\Bigr\}
\tilde{\tau}[\epsilon(\boldq)]}
{\sum\limits_{\boldq}\tilde{e}^{z}(\boldq)^{2}
\Bigl\{-\frac{\partial n(\epsilon_{\boldq})}{\partial \epsilon(\boldq)}\Bigr\}
\tilde{\tau}[\epsilon(\boldq)]}.
\end{align} 
Since the contributions for small $q=|\boldq|$ are dominant 
due to the $\{-\frac{\partial n[\epsilon(\boldq)]}{\partial \epsilon(\boldq)}\}$, 
we estimate $\kappa_{xx}/\kappa_{zz}$ 
by replacing $\boldq$ in $\tilde{e}^{x}(\boldq)^{2}$ and $\tilde{e}^{z}(\boldq)^{2}$ 
by $\boldq_{0}$; 
as described above, 
$\boldq_{0}$ is a certain momentum whose magnitude is small. 
As a result, 
$\kappa_{xx}/\kappa_{zz}$ is estimated as follows:
\begin{align}
\frac{\kappa_{xx}}{\kappa_{zz}}
\sim 
\frac{\tilde{e}^{x}(\boldq_{0})^{2}}{\tilde{e}^{z}(\boldq_{0})^{2}}
=\frac{e_{11}^{x}(\boldq_{0})^{2}+e_{12}^{x}(\boldq_{0})^{2}\sinh^{2}2\theta_{\boldq_{0}}}
{e_{11}^{z}(\boldq_{0})^{2}+e_{12}^{z}(\boldq_{0})^{2}\sinh^{2}2\theta_{\boldq_{0}}}.\label{eq:anisotropy}
\end{align}
To estimate this quantity, 
we calculate the numerator and denominator
by considering the dominant terms including the leading correction from $D/J$
because $D$ will be typically smaller than $J$. 
After some calculations, 
described in Appendix H, 
we obtain
\begin{align}
\frac{\kappa_{xx}}{\kappa_{zz}}
\sim 1+\frac{3}{8}\Bigl(\frac{D}{J}\Bigr)^{2}q_{0}^{2}.\label{eq:anisotropy-final}
\end{align}
Thus 
the inplane anisotropy of longitudinal thermal conductivities 
is proportional to the squared ratio of the Dzyaloshinsky-Moriya interaction 
to the Heisenberg interaction: 
\begin{align}
\frac{\kappa_{xx}-\kappa_{zz}}{\kappa_{zz}}\propto \Bigl(\frac{D}{J}\Bigr)^{2}.
\end{align}
This result indicates that 
it is possible to estimate the magnitude of $D/J$ by measuring the inplane anisotropy 
of longitudinal thermal conductivities.

\section{Discussion}

We first compare properties for the disordered spiral magnet 
and the disordered collinear antiferromagnet~\cite{Lett}. 
The same properties are global time-reversal symmetry, 
the diverging behavior of the particle-particle type four-point vertex function 
for the back scattering, 
and the negative logarithmic divergence of $\Delta \kappa_{\alpha\alpha}$ for $L\rightarrow \infty$. 
This suggests that 
the weak localization of magnons is not unique
only for the disordered collinear antiferromagnet, 
but ubiquitous for the disordered magnets having global time-reversal symmetry. 
The major differences are the alignments of spins 
and the inplane anisotropy of longitudinal thermal conductivities: 
only one of the three components of
$\langle \hat{S}^{\alpha}_{\boldi}\rangle$ ($\alpha=x,y,z$) 
is finite in the collinear antiferromagnet, 
whereas two are finite in the screw-type spiral magnet; 
$\kappa_{xx}$ and $\kappa_{yy}$ are the same
in the collinear antiferromagnet on a $xy$ plane, 
whereas $\kappa_{xx}$ and $\kappa_{zz}$ are different
in the screw-type spiral magnet 
on a $xz$ plane. 
The difference in the spin alignments 
arises from the effect of the Dzyaloshinsky-Moriya interaction, 
which is finite only for the screw-type spiral magnet. 
The difference in the inplane anisotropy of $\kappa_{\alpha\alpha}$ 
arises from the different spin alignments; 
in the screw-type spiral magnet 
the inplane longitudinal thermal conductivities become anisotropic 
due to the difference between magnon propagations  
parallel and perpendicular to the spiral axis. 

We next discuss the validity of our approximation. 
We used the linear-spin-wave approximation, 
which took account of the quadratic terms of magnon operators. 
We believe this approximation is sufficient to analyze transport properties 
for low-energy magnons in two-dimensional magnets at low temperatures 
because 
several previous theoretical studies suggest that 
the terms neglected in the linear-spin-wave approximation 
may not change our main results at least qualitatively. 
First, 
some studies~\cite{RMP-AF} for a $S=1/2$ Heisenberg antiferromagnet on a square lattice  
show that 
the effects of the zero-point fluctuations and the magnon-magnon interaction 
are small at low temperatures. 
This result suggests that 
the corrections due to 
the zeroth-order term of magnon operators and the fourth-order (and higher-order) terms 
will be small at least at low temperatures. 
Then the theoretical studies for noncollinear antiferromagnets~\cite{Shiba-chiral,RMP-chiral} 
show that 
the third-order terms of magnon operators induce the magnon-magnon interaction 
characteristic of noncollinear magnets, such as spiral magnets, 
and that its effects on the energy dispersion and damping for low-energy magnons 
are small. 
Since low-energy magnons give the dominant contributions to 
the longitudinal thermal conductivities of magnons, 
the third-order terms also may not change our main results at least qualitatively. 

We turn to implications for further theoretical studies. 
First, 
an analogy with the disordered collinear antiferromagnet~\cite{Full} suggests that 
the disordered screw-type spiral magnet 
may show a characteristic property of magnetothermal magnon transport 
in the presence of a weak external magnetic field. 
This could be demonstrated by the theory 
for the disordered spiral magnet with the weak external magnetic field. 
Second, 
by combining our theory without using the two simplifications 
with first-principles calculations, 
it is possible to study material varieties of the weak localization of magnons 
in various disordered magnets. 
For the first-principles calculations, 
a set of Eqs. (\ref{eq:kappa-WL}){--}(\ref{eq:Pi}) is more appropriate than 
the theory using the two simplifications. 
Third, 
our theory can be extended to not only other disordered spiral magnets, 
but also disordered chiral magnets, which have finite spin scalar chirality. 
Whereas spin scalar chirality for certain three sites breaks 
local time-reversal symmetry for the three sites, 
global time-reversal symmetry could hold 
in some disordered chiral magnets; 
this could be possible 
if the disordered chiral magnet has the magnetic structure 
consisting of time-reversal symmetric pairs for spin scalar chirality 
[e.g., $\langle \hat{\boldS}_{\boldi}\cdot(\hat{\boldS}_{\boldj}\times \hat{\boldS}_{\boldk})\rangle$ and
  $\langle \hat{\boldS}_{\boldi^{\prime}}\cdot(\hat{\boldS}_{\boldj^{\prime}}\times \hat{\boldS}_{\boldk^{\prime}})\rangle=-\langle \hat{\boldS}_{\boldi}\cdot (\hat{\boldS}_{\boldj}\times \hat{\boldS}_{\boldk})\rangle$]. 

We finally discuss implications for experiments. 
First, 
the weak localization of magnons in our disordered spiral magnet can be experimentally observed 
by measuring $\kappa_{\alpha\alpha}$. 
If $\kappa_{\alpha\alpha}$ is measured at very low temperatures, 
at which the inelastic scattering due to the magnon-magnon interaction 
is negligible, 
the weak localization of magnons will be observed as the drastic suppression 
of the magnon thermal current parallel to temperature gradient 
as a result of the logarithmic dependence of $\Delta\kappa_{\alpha\alpha}$ on $L$. 
If the measurement is performed at low temperatures, 
at which the inelastic scattering is small but non-negligible, 
the weak localization of magnons would be observed as 
the logarithmic temperature dependence of $\kappa_{\alpha\alpha}$; 
this is based on a similar argument~\cite{Lett} to the effect of 
the inelastic scattering for electrons~\cite{Thouless-Inela,Anderson-Inela}. 
Second, the inplane anisotropy of longitudinal thermal conductivities 
could be used to experimentally estimate the magnitude of $D/J$ 
in the screw-type spiral magnet. 
In particular, 
this method may be convenient for experimentally estimating 
whether the Dzyaloshinsky-Moriya interaction is small or large. 
Third,  
our main results, 
the weak localization of magnons and 
the inplane anisotropy of longitudinal thermal conductivities, 
may be realized in a real material, for example, 
Ba$_{2}$Cu$_{1-x}$Ag$_{x}$Ge$_{2}$O$_{7}$. 
The magnetic properties of Ba$_{2}$CuGe$_{2}$O$_{7}$ 
are described by the spin Hamiltonian consisting of 
the antiferromagnetic Heisenberg interaction and the Dzyaloshinsky-Moriya interaction 
for $S=1/2$ Cu$^{2+}$ ions~\cite{CuSpiral1,CuSpiral2,CuSpiral3}. 
Furthermore, 
Ba$_{2}$CuGe$_{2}$O$_{7}$ at very low temperatures can be regarded as 
a screw-type spiral magnet on a square lattice~\cite{CuSpiral1,CuSpiral2,CuSpiral3}; 
in this magnet, 
the spin alignment along a direction on the square lattice is spiral, 
and the spin alignment along the perpendicular direction is antiferromagnetic. 
These spin alignments are similar to those for our screw-type spiral magnet, 
whereas there are some differences in the details. 
Then 
replacing part of Cu ions by Ag ions 
will be suitable for impurities 
because this replacement keeps the spin quantum number $S$ unchanged 
and its main effect is to modify the exchange interactions~\cite{Lett,Full}. 
We believe 
Ba$_{2}$Cu$_{1-x}$Ag$_{x}$Ge$_{2}$O$_{7}$ is a probable material 
for the weak localization of magnons and 
the inplane anisotropy of longitudinal thermal conductivities. 
This is 
because the inplane anisotropy results from the difference 
between magnon propagation along the spiral spin alignment and 
along the antiferromagnetic spin alignment, 
and because the magnetic structure of Ba$_{2}$Cu$_{1-x}$Ag$_{x}$Ge$_{2}$O$_{7}$ 
without external fields may have global time-reversal symmetry 
and such spin alignments.

\section{Summary}

We have studied the longitudinal thermal conductivities of magnons 
in the disordered screw-type spiral magnet in the weak-localization regime. 
We used the spin Hamiltonian consisting of 
the antiferromagnetic Heisenberg interaction 
and the Dzyaloshinsky-Moriya interaction 
on a square lattice on a $xz$ plane. 
We also considered the mean-field type spin Hamiltonian of impurities 
by treating disorder effects as 
the changes of these exchange interactions. 
By using the linear-response theory with 
the linear-spin-wave approximation for the screw-type spiral magnet 
and performing the perturbation calculations, 
we derived the longitudinal thermal conductivities including the main correction term 
in the weak-localization regime. 
We showed that 
$\kappa_{xx}$ and $\kappa_{zz}$ are different due to the difference 
between magnon propagations parallel and perpendicular to the spiral axis. 
This anisotropy is different from the isotropic result 
in the disordered two-dimensional antiferromagnet, and 
its measurement may be useful 
for experimentally estimating the magnitude of $D/J$. 
We also showed that 
the main correction term gives the negative logarithmic divergence 
in the thermodynamic limit 
due to the critical back scattering. 
This is the same as the weak localization of magnons 
in the disordered two-dimensional antiferromagnet~\cite{Lett}, 
and thus suggests the generality of the weak localization of magnons 
in the disordered two-dimensional magnets having global time-reversal symmetry. 

\begin{acknowledgments}
This work was supported by CREST, JST, and Grant-in-Aid 
for Scientific Research (A) (17H01052) from MEXT, Japan.
\end{acknowledgments}

\appendix
\begin{widetext}
\section{Remarks about coherence of the back scattering}

\begin{figure*}[tb]
\includegraphics[width=150mm]{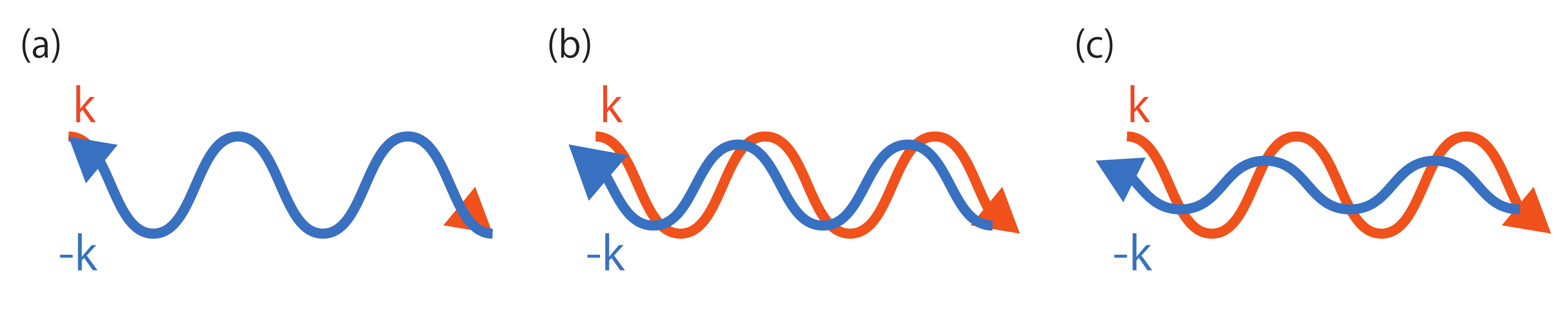}
\caption{
Schematic illustrations of the possible relations between 
the wave packets for $\boldk$ and $-\boldk$. 
Panel (a) shows the perfectly coherent case, 
whereas panels (b) and (c) show imperfectly coherent cases; 
in our definition 
the back scattering only in panel (a) is the coherent one. 
The amplitude difference and phase difference in panel (b) 
are smaller than in panel (c).} 
\label{fig4}
\end{figure*}
In this appendix 
we explain our definition of the coherent back scattering and its implications. 
We also show the relation between the coherence of the back scattering 
and the time-reversal symmetry of a system. 

In our definition the back scattering is coherent 
only if the amplitude and the phase of the back scattered wave packet, 
the wave packet for $-\boldk$, 
are the same as those of the wave packet for $\boldk$ [Fig. \ref{fig4}(a)]. 
Even if the amplitude remains high and 
the phase difference is small [Fig. \ref{fig4}(b)], 
such back scattering is not the coherent one in our definition. 
Figure \ref{fig4} shows 
the possible relations between the wave packets for $\boldk$ and $-\boldk$.

We have used this definition because 
only in the perfectly coherent case 
the wave packets for $\boldk$ and $-\boldk$ can form 
the standing wave even in the weak-localization regime. 
In an imperfectly coherent case 
the imbalance between the wave packets for $\boldk$ and $-\boldk$ 
could lead to finite conduction in the weak-localization regime. 
Actually, in such a case the back scattering amplitude is suppressed~\cite{PhotoLoc} 
compared with that in the perfectly coherent case. 
This suppression, as well as the suppression of the critical back scattering~\cite{Full}, 
results from the effect of the time-reversal symmetry breaking. 

Then we can see the relation between coherence of the back scattering and 
time-reversal symmetry of a system from the following arguments. 
To see that relation, 
we argue the property of time reversal for a single-particle Green's function. 
The retarded single-particle Green's function is defined as
$G^{(\textrm{R})}(\boldk,\omega)\equiv
\langle \boldk |\frac{1}{\omega-H+i\delta}|\boldk\rangle$, 
where $H$ is the Hamiltonian of a system. 
We assume that $H$ has time-reversal symmetry. 
This means $H\Theta=\Theta H$, 
where $\Theta$ is the time-reversal operator~\cite{Sakurai}. 
Since $\Theta$ is antiunitary, 
we have
$\langle \beta |\alpha\rangle=\langle \alpha^{\prime} |\beta^{\prime}\rangle$, 
where $|\alpha^{\prime}\rangle=\Theta|\alpha\rangle$ 
and $|\beta^{\prime}\rangle=\Theta|\beta\rangle$.  
By applying this equation to the case 
for $|\alpha\rangle=\frac{1}{\omega-H+i\delta}|\boldk\rangle$ 
and $|\beta\rangle=|\boldk\rangle$ and using $H\Theta=\Theta H$, 
we obtain 
$G^{(\textrm{R})}(\boldk,\omega)=G^{(\textrm{R})}(-\boldk,\omega)$. 
This equation shows that as a result of time-reversal symmetry 
the single particles for $\boldk$ and $-\boldk$ 
have the same amplitude and phase. 

\section{Ground-state properties of $\hat{H}_{0}$} 
In this appendix 
we show the ground-state properties of $\hat{H}_{0}$ 
in the mean-field approximation. 
For the details of the mean-field approximation for a spin Hamiltonian, 
see, for example, Refs. \onlinecite{NA-Ir}, \onlinecite{NA-pyro}, and \onlinecite{Yosida-text}.
In the mean-field approximation 
we can determine the most stable ground state 
for Eq. (\ref{eq:H0-spin})
by finding the lowest eigenvalue and the eigenfunction 
for the following equation under the hard-spin constraint:
\begin{align}
\langle \hat{H}_{0}\rangle 
=&\sum\limits_{\boldi,\boldj}\sum\limits_{\alpha,\beta=x,y,z}
M_{\alpha \beta}(\boldi,\boldj)
\langle \hat{S}_{\boldi}^{\alpha}\rangle
\langle \hat{S}_{\boldj}^{\beta}\rangle\notag\\
=&\sum\limits_{\boldq}\sum\limits_{\alpha,\beta=x,y,z}
M_{\alpha \beta}(\boldq)
\langle \hat{S}_{\boldq}^{\alpha}\rangle^{\ast}
\langle \hat{S}_{\boldq}^{\beta}\rangle\notag\\
=&\sum\limits_{\boldq}
\langle \hat{\boldS}_{\boldq}\rangle^{\dagger}
\left(
\begin{array}{@{\,}ccc@{\,}}
J(\boldq) & 0 & D(\boldq)^{\ast}\\
0 & J(\boldq) & 0\\
D(\boldq) & 0 & J(\boldq)
\end{array}
\right)
\langle \hat{\boldS}_{\boldq}\rangle,
\end{align}
where 
\begin{align}
&\langle \hat{S}_{\boldi}^{\alpha}\rangle
=\frac{1}{\sqrt{N}}
\sum\limits_{\boldq}
e^{-i\boldq\cdot \boldi}\langle \hat{S}_{\boldq}^{\alpha}\rangle,\\
&M_{\alpha\alpha}(\boldq)=J(\boldq)=J(\cos q_{x}+\cos q_{z}),\label{eq:Jq}\\ 
&M_{zx}(\boldq)=M_{xz}(\boldq)^{\ast}=D(\boldq)=iD\sin q_{x},\label{eq:Dq}
\end{align} 
and 
$\langle \hat{S}_{\boldi}^{\alpha}\rangle$ satisfies the hard-spin constraint, 
\begin{align}
S^{2}=\frac{1}{N}\sum\limits_{\boldi}\sum\limits_{\alpha}
|\langle \hat{S}_{\boldi}^{\alpha}\rangle|^{2}.\label{eq:HardS}
\end{align}
Since the eigenvalues of the $3\times 3$ matrix $M_{\alpha\beta}(\boldq)$ are 
$\lambda_{0}(\boldq)=J(\boldq)$, 
$\lambda_{+}(\boldq)=J(\boldq)+ |D(\boldq)|$, 
and $\lambda_{-}(\boldq)=J(\boldq)- |D(\boldq)|$, 
the minimum of $\lambda_{-}(\boldq)$ is the lowest eigenvalue. 
For finite $J$ and $D$, 
$\lambda_{-}(\boldq)$ is minimum at $\boldq=\boldQ={}^{t}(Q_{x}\ Q_{z})$, 
where $Q_{z}=\pi$, and $Q_{x}$ is determined by 
\begin{align}
\cos Q_{x}=-\frac{J}{\sqrt{J^{2}+D^{2}}},\ 
\sin Q_{x}=\frac{D}{\sqrt{J^{2}+D^{2}}}.\label{eq:cosQ,sinQ}
\end{align}
Since $\lambda_{-}(\boldQ)=-J-\sqrt{J^{2}+D^{2}}$ 
is smaller than $\lambda_{-}(\boldq)$ at $\boldq=\boldQ_{\textrm{AF}}={}^{t}(\pi\ \pi)$, 
$\lambda_{-}(\boldQ_{\textrm{AF}})=-2J$, 
the magnetic state for $\boldq=\boldQ$ is more stable 
than the antiferromagnetic state even for tiny $D$. 
Then
we can determine the eigenfunction for the most stable ground state as follows. 
In the mean-field approximation for the magnetic state for $\boldq=\boldQ$ 
only $\langle \hat{\boldS}_{\boldQ}\rangle$ and $\langle \hat{\boldS}_{-\boldQ}\rangle$ 
are finite 
and the other $\langle \hat{\boldS}_{\boldq}\rangle$'s are zero. 
Since $\langle \hat{\boldS}_{\boldQ}\rangle$ is given by the eigenfunction 
for $M_{\alpha\beta}(\boldQ)$, 
$\langle \hat{\boldS}_{\boldQ}\rangle=A{}^{t}(i\ 0\ 1)$, 
and $A$ is determined from Eq. (\ref{eq:HardS}), 
the magnetic structure for $\boldq=\boldQ$ is described by
\begin{align}
\langle \boldS_{\boldi}\rangle
=S
\left(
\begin{array}{@{\,}c@{\,}}
\sin \boldQ\cdot\boldi\\
0\\
\cos \boldQ\cdot\boldi
\end{array}
\right).
\end{align} 
This equation with Eq. (\ref{eq:cosQ,sinQ}) and $Q_{z}=\pi$ show that 
the alignment of spins in a $x$ direction is spiral 
and the alignment in a $z$ direction is antiferromagnetic 
(see Fig. \ref{fig1}).  

\section{Derivation of Eq. (\ref{eq:H0-S'})}

In this appendix we derive Eq. (\ref{eq:H0-S'}).
By using Eqs. (\ref{eq:Sx'}){--}(\ref{eq:Sz'}), 
we can express Eq. (\ref{eq:H0-spin}) as follows:
\begin{align}
\hat{H}_{0}
=
&\sum\limits_{\langle\boldi,\boldj\rangle}
\Bigl\{J_{\boldi\boldj}\cos[\boldQ\cdot(\boldi-\boldj)]
+D_{\boldi\boldj}\sin[\boldQ\cdot(\boldi-\boldj)]\Bigr\}
(\hat{S}_{\boldi}^{\prime x}\hat{S}_{\boldj}^{\prime x}
+\hat{S}_{\boldi}^{\prime z}\hat{S}_{\boldj}^{\prime z})
+\sum\limits_{\langle\boldi,\boldj\rangle}
J_{\boldi\boldj}
\hat{S}_{\boldi}^{\prime y}\hat{S}_{\boldj}^{\prime y}\notag\\
&+\sum\limits_{\langle\boldi,\boldj\rangle}
\Bigl\{J_{\boldi\boldj}\sin[\boldQ\cdot(\boldi-\boldj)]
-D_{\boldi\boldj}\cos[\boldQ\cdot(\boldi-\boldj)]
\Bigr\}
(\hat{S}_{\boldi}^{\prime z}\hat{S}_{\boldj}^{\prime x}
-\hat{S}_{\boldi}^{\prime x}\hat{S}_{\boldj}^{\prime z}).\label{eq:H0-S'-pre}
\end{align}
By using Eqs. (\ref{eq:J}), (\ref{eq:DM}), and (\ref{eq:cosQ,sinQ}) 
and $Q_{z}=\pi$, 
the coefficients of the first and third terms in the above equation 
can be rewritten in a simpler expression: 
the coefficients for $\boldj-\boldi=(1\ 0)$ are
\begin{align}
&J_{\boldi\boldj}\cos[\boldQ\cdot(\boldi-\boldj)]
+D_{\boldi\boldj}\sin[\boldQ\cdot(\boldi-\boldj)]=-\sqrt{J^{2}+D^{2}},\\
&J_{\boldi\boldj}\sin[\boldQ\cdot(\boldi-\boldj)]
-D_{\boldi\boldj}\cos[\boldQ\cdot(\boldi-\boldj)]=0,
\end{align}
and the coefficients for $\boldj-\boldi=(0\ 1)$ are
\begin{align}
&J_{\boldi\boldj}\cos[\boldQ\cdot(\boldi-\boldj)]
+D_{\boldi\boldj}\sin[\boldQ\cdot(\boldi-\boldj)]=-J,\\
&J_{\boldi\boldj}\sin[\boldQ\cdot(\boldi-\boldj)]
-D_{\boldi\boldj}\cos[\boldQ\cdot(\boldi-\boldj)]=0.
\end{align}
Thus Eq. (\ref{eq:H0-S'-pre}) is reduced to Eq. (\ref{eq:H0-S'}).

\section{Properties of the energy dispersion relation of magnon bands}

In this appendix 
we explain several important properties 
of the energy dispersion relation of magnon bands for our spiral magnet. 
Before explaining the properties, 
we show the equation of $\epsilon(\boldq)$ 
in terms of $J$ and $D$. 
Since $A(\boldq)$ and $B(\boldq)$ for our model 
are expressed as
\begin{align}
A(\boldq)=S(\sqrt{J^{2}+D^{2}}+J)-\frac{S}{2}(\sqrt{J^{2}+D^{2}}-J)\cos q_{x}\label{eq:Aq}
\end{align}
and 
\begin{align}
B(\boldq)=-\frac{S}{2}(\sqrt{J^{2}+D^{2}}+J)\cos q_{x}-SJ\cos q_{z},\label{eq:Bq}
\end{align}
respectively, 
we obtain
\begin{align}
\epsilon(\boldq)
=&2\sqrt{A(\boldq)^{2}-B(\boldq)^{2}}\notag\\
=&2S\Bigl[
2J^{2}+D^{2}+2J\sqrt{J^{2}+D^{2}}-D^{2}\cos q_{x}
-(J^{2}+J\sqrt{J^{2}+D^{2}})\cos q_{x}\cos q_{z}\notag\\
&\ \ \ \ \ -J\sqrt{J^{2}+D^{2}}\cos^{2} q_{x}-J^{2}\cos^{2}q_{z}
\Bigr]^{\frac{1}{2}}.\label{eq:e-q}
\end{align}

If we set $\boldq=\boldzero$ in Eq. (\ref{eq:e-q}), 
we obtain $\epsilon(\boldzero)=0$. 
This means that the screw-type spiral magnet has the Goldstone-type gapless excitation. 
This is consistent with the argument based on the rotational symmetry 
in the spin space~\cite{NA-JS} 
because in our spiral magnet 
two of the three components of $\langle \hat{S}_{\boldi}^{\alpha}\rangle$ 
(i.e., $\langle \hat{S}_{\boldi}^{x}\rangle$ 
and $\langle \hat{S}_{\boldi}^{z}\rangle$) are finite 
and because 
in such a case 
the Goldstone-type gapless excitation is expected to exist without magnetic anisotropy terms. 

Then, 
since the magnon energy is non-negative, 
the magnon energy is minimum at $\boldq=\boldzero$ in our spiral magnet. 
This result is consistent with the assumption that 
the screw-type spiral state remains stable even including low-energy excitations, i.e., magnons, 
because magnons describe the displacement of spins from the ground-state alignment, 
because the magnon for $\boldq=\boldzero$ corresponds to the uniform displacement, 
and because the uniform displacement 
induces no additional symmetry breaking.  
If the magnon energy is minimum at $\boldq=\boldQ_{\textrm{I}}$ and $-\boldQ_{\textrm{I}}$, 
this means either that 
magnons break a certain inversion symmetry which exists without magnons, 
or that it is necessary to choose a more stable ground state as the starting point 
for considering magnons. 

\section{Derivation of Eq. (\ref{eq:JE})}
In this appendix 
we derive Eq. (\ref{eq:JE}) from Eq. (\ref{eq:JE-def}) for $\hat{H}=\hat{H}_{0}$ 
for Eq. (\ref{eq:H0-mag-site}). 
This derivation consists of four steps. 
First, 
we decompose $\hat{h}_{\boldi}$ and $\hat{h}_{\boldj}$ in Eq. (\ref{eq:JE-def}) as follows: 
$\hat{h}_{\boldi}=\hat{A}_{\boldi}+\hat{A}_{\boldi}^{\dagger}$ 
and $\hat{h}_{\boldj}=\hat{A}_{\boldj}+\hat{A}_{\boldj}^{\dagger}$, 
where 
\begin{align}
\hat{A}_{\boldi}
=\frac{S}{4}\sum\limits_{\boldl}M_{\boldi \boldl}\hat{b}_{\boldi}^{\dagger}\hat{b}_{\boldl}
-\frac{S}{4}\sum\limits_{\boldl}\tilde{J}^{(+)}_{\boldi\boldl}\hat{b}_{\boldi}\hat{b}_{\boldl},
\label{eq:Ai}
\end{align}
with $M_{\boldi\boldl}=2\sum_{\boldk}\tilde{J}_{\boldi\boldk}\delta_{\boldi,\boldl}-\tilde{J}_{\boldi\boldl}^{(-)}$. 
Because of these decompositions, 
Eq. (\ref{eq:JE-def}) is reduced to
\begin{align}
\hat{\boldJ}_{E}
=&i\sum\limits_{\boldm,\boldn}\boldr_{\boldn}[\hat{A}_{\boldm},\hat{A}_{\boldn}]
+\Bigl(
i\sum\limits_{\boldm,\boldn}\boldr_{\boldn}[\hat{A}_{\boldm},\hat{A}_{\boldn}]
\Bigr)^{\dagger}
+i\sum\limits_{\boldm,\boldn}\boldr_{\boldn}[\hat{A}_{\boldm},\hat{A}_{\boldn}^{\dagger}]
+\Bigl(
i\sum\limits_{\boldm,\boldn}\boldr_{\boldn}[\hat{A}_{\boldm},\hat{A}_{\boldn}^{\dagger}]
\Bigr)^{\dagger}.\label{eq:JE-second}
\end{align}
Second, 
we calculate $[\hat{A}_{\boldm},\hat{A}_{\boldn}]$ and 
$[\hat{A}_{\boldm},\hat{A}_{\boldn}^{\dagger}]$. 
The results are as follows: 
\begin{align}
[\hat{A}_{\boldm},\hat{A}_{\boldn}]
&=\Bigl(\frac{S}{4}\Bigr)^{2}
\sum\limits_{\boldj,\boldl}M_{\boldm\boldj}M_{\boldn\boldl}
(\hat{b}_{\boldm}^{\dagger}\hat{b}_{\boldl}\delta_{\boldj,\boldn}
-\hat{b}_{\boldn}^{\dagger}\hat{b}_{\boldj}\delta_{\boldl,\boldm})
+
\Bigl(\frac{S}{4}\Bigr)^{2}
\sum\limits_{\boldj,\boldl}M_{\boldm\boldj}\tilde{J}^{(+)}_{\boldn\boldl}
(\hat{b}_{\boldn}\hat{b}_{\boldj}\delta_{\boldl,\boldm}
+\hat{b}_{\boldl}\hat{b}_{\boldj}\delta_{\boldn,\boldm})\notag\\
&-
\Bigl(\frac{S}{4}\Bigr)^{2}
\sum\limits_{\boldj,\boldl}\tilde{J}^{(+)}_{\boldm\boldj}M_{\boldn\boldl}
(\hat{b}_{\boldm}\hat{b}_{\boldl}\delta_{\boldj,\boldn}
+\hat{b}_{\boldj}\hat{b}_{\boldl}\delta_{\boldn,\boldm}),\\
[\hat{A}_{\boldm},\hat{A}_{\boldn}^{\dagger}]
&=\Bigl(\frac{S}{4}\Bigr)^{2}
\sum\limits_{\boldj,\boldl}M_{\boldm\boldj}M_{\boldn\boldl}
(\hat{b}_{\boldm}^{\dagger}\hat{b}_{\boldn}\delta_{\boldj,\boldl}
-\hat{b}_{\boldl}^{\dagger}\hat{b}_{\boldj}\delta_{\boldn,\boldm})
-
\Bigl(\frac{S}{4}\Bigr)^{2}
\sum\limits_{\boldj,\boldl}M_{\boldm\boldj}\tilde{J}^{(+)}_{\boldn\boldl}
(\hat{b}_{\boldm}^{\dagger}\hat{b}_{\boldn}^{\dagger}\delta_{\boldl,\boldj}
+\hat{b}_{\boldm}^{\dagger}\hat{b}_{\boldl}^{\dagger}\delta_{\boldn,\boldj})\notag\\
&-
\Bigl(\frac{S}{4}\Bigr)^{2}
\sum\limits_{\boldj,\boldl}\tilde{J}^{(+)}_{\boldm\boldj}M_{\boldn\boldl}
(\hat{b}_{\boldm}\hat{b}_{\boldn}\delta_{\boldj,\boldl}
+\hat{b}_{\boldj}\hat{b}_{\boldn}\delta_{\boldl,\boldm})\notag\\
&+
\Bigl(\frac{S}{4}\Bigr)^{2}
\sum\limits_{\boldj,\boldl}\tilde{J}^{(+)}_{\boldm\boldj}\tilde{J}^{(+)}_{\boldn\boldl}
(\hat{b}_{\boldm}\hat{b}_{\boldn}^{\dagger}\delta_{\boldj,\boldl}
+\hat{b}_{\boldm}\hat{b}_{\boldl}^{\dagger}\delta_{\boldj,\boldn}
+\hat{b}_{\boldl}^{\dagger}\hat{b}_{\boldj}\delta_{\boldm,\boldn}
+\hat{b}_{\boldn}^{\dagger}\hat{b}_{\boldj}\delta_{\boldl,\boldm}).
\end{align}
Third, 
we combine these equations with Eq. (\ref{eq:JE-second}). 
After some algebra, 
we obtain
\begin{align}
\hat{\boldJ}_{E}
=&2i\Bigl(\frac{S}{4}\Bigr)^{2}\sum\limits_{\boldm,\boldn,\boldl}
(-\boldr_{\boldn}+\boldr_{\boldl})M_{\boldm\boldl}M_{\boldn\boldm}
\hat{b}_{\boldn}^{\dagger}\hat{b}_{\boldl}
+2i\Bigl(\frac{S}{4}\Bigr)^{2}\sum\limits_{\boldm,\boldn,\boldl}
(\boldr_{\boldn}-\boldr_{\boldm})\tilde{J}^{(+)}_{\boldm\boldl}\tilde{J}^{(+)}_{\boldn\boldl}
\hat{b}_{\boldm}\hat{b}_{\boldn}^{\dagger}\notag\\
&+2i\Bigl(\frac{S}{4}\Bigr)^{2}\sum\limits_{\boldm,\boldn,\boldl}
(\boldr_{\boldn}-\boldr_{\boldl})\tilde{J}^{(+)}_{\boldn\boldm}M_{\boldm\boldl}
\hat{b}_{\boldn}\hat{b}_{\boldl}
-2i\Bigl(\frac{S}{4}\Bigr)^{2}\sum\limits_{\boldm,\boldn,\boldl}
(\boldr_{\boldn}-\boldr_{\boldl})M_{\boldm\boldl}\tilde{J}^{(+)}_{\boldn\boldm}
\hat{b}_{\boldl}^{\dagger}\hat{b}_{\boldn}^{\dagger}.\label{eq:JE-third}
\end{align}
Fourth, 
by using the Fourier coefficient of each quantity in Eq. (\ref{eq:JE-third}), 
we express $\hat{\boldJ}_{E}$ as a function of a momentum. 
By carrying out this calculation, 
we obtain Eq. (\ref{eq:JE}).

\section{Derivation of Eq. (\ref{eq:Pi-approx})}
In this appendix 
we derive Eq. (\ref{eq:Pi-approx}) from Eq. (\ref{eq:Pi}) with 
Eqs. (\ref{eq:D^R}) and (\ref{eq:D^A}). 
We here describe this derivation only for $\epsilon > 0$ 
because the expression for $\epsilon < 0$ can be similarly derived. 
By substituting Eqs. (\ref{eq:D^R}) and (\ref{eq:D^A}) for $\epsilon > 0$
into Eq. (\ref{eq:Pi}), 
we can express $\Pi_{abcd}(\boldQ,\epsilon)$ for $\epsilon > 0$ as follows:
\begin{align}
\Pi_{abcd}(\boldQ,\epsilon)
=\sum\limits_{\boldq_{1}}
\frac{U_{b\alpha}(\boldq_{1})U_{c\alpha}(\boldq_{1})
U_{d\alpha}(\boldQ-\boldq_{1})U_{a\alpha}(\boldQ-\boldq_{1})}
{[\epsilon-\epsilon(\boldq_{1})+i\tilde{\gamma}(\epsilon)]
[\epsilon-\epsilon(\boldQ-\boldq_{1})-i\tilde{\gamma}(\epsilon)]}.\label{eq:Pi-1st}
\end{align}
Since for small $Q$, $U_{a\nu}(\boldQ-\boldq_{1})\sim U_{a\nu}(\boldq_{1})$ 
and $\epsilon(\boldQ-\boldq_{1})\sim 
\epsilon(\boldq_{1})-\frac{\partial \epsilon(\boldq_{1})}{\partial \boldq_{1}}\cdot \boldQ
=\epsilon(\boldq_{1})-\boldv_{\boldq_{1}}\cdot \boldQ$, 
we can approximate Eq. (\ref{eq:Pi-1st}) as follows: 
\begin{align}
\Pi_{abcd}(\boldQ,\epsilon)
\sim\sum\limits_{\boldq_{1}}
\frac{U_{b\alpha}(\boldq_{1})U_{c\alpha}(\boldq_{1})
U_{d\alpha}(\boldq_{1})U_{a\alpha}(\boldq_{1})}
{[\epsilon-\epsilon(\boldq_{1})+i\tilde{\gamma}(\epsilon)]
[\epsilon-\epsilon(\boldq_{1})+\boldv_{\boldq_{1}}\cdot \boldQ-i\tilde{\gamma}(\epsilon)]}.
\label{eq:Pi-2nd}
\end{align}
For a rough estimate of Eq. (\ref{eq:Pi-2nd}) 
we approximate momentum-dependent $U_{a\alpha}(\boldq_{1})$ and $\boldv_{\boldq_{1}}$ 
as the typical values at a certain, small momentum $\boldq_{0}$, 
$U_{a\alpha}(\boldq_{0})=u_{a\alpha}$ and $\boldv_{\boldq_{0}}$; 
this will be sufficient because the dominant contributions come from 
the small-$q_{1}$ contributions. 
As a result of this approximation, 
Eq. (\ref{eq:Pi-2nd}) is expressed as follows:
\begin{align}
&\Pi_{abcd}(\boldQ,\epsilon)
\sim 
\sum\limits_{\boldq_{1}}
\frac{u_{b\alpha}u_{c\alpha}u_{d\alpha}u_{a\alpha}}
{[\epsilon-\epsilon(\boldq_{1})+i\tilde{\gamma}(\epsilon)]
[\epsilon-\epsilon(\boldq_{1})+\boldv_{\boldq_{0}}\cdot \boldQ-i\tilde{\gamma}(\epsilon)]}.
\label{eq:Pi-3rd}
\end{align}
Here we have replaced $(\cosh^{2}\theta_{\boldq_{1}}+\sinh^{2}\theta_{\boldq_{1}})^{2}$ 
in $\tilde{\gamma}(\epsilon)$ 
by $(\cosh^{2}\theta_{\boldq_{0}}+\sinh^{2}\theta_{\boldq_{0}})^{2}=(c_{0}^{2}+s_{0}^{2})^{2}$. 
Then, 
by replacing the summation over $\boldq_{1}$ by the corresponding integral 
and carrying out this integral, 
we obtain  
\begin{align}
\Pi_{abcd}(\boldQ,\epsilon)
&\sim 
u_{b\alpha}u_{c\alpha}u_{d\alpha}u_{a\alpha}
N\pi\rho(\epsilon)\tilde{\tau}(\epsilon)
[1-\frac{1}{8}\boldv_{\boldq_{0}}^{2}Q^{2}\tilde{\tau}(\epsilon)^{2}]
= \frac{u_{b\alpha}u_{c\alpha}u_{d\alpha}u_{a\alpha}}{\gamma_{\textrm{imp}}(c_{0}^{2}+s_{0}^{2})^{2}}
[1-D_{\textrm{S}}(\epsilon)Q^{2}\tilde{\tau}(\epsilon)].
\end{align}
This is Eq. (\ref{eq:Pi-approx}) for $\epsilon > 0$. 
We can also obtain Eq. (\ref{eq:Pi-approx}) for $\epsilon < 0$ 
by using Eqs. (\ref{eq:D^R}) and (\ref{eq:D^A}) for $\epsilon < 0$ 
and carrying out the similar calculation. 

\section{Derivation of Eq. (\ref{eq:Gamma-approx})}
In this appendix 
we derive Eq. (\ref{eq:Gamma-approx}). 
This derivation consists of three steps. 
First, 
we rewrite the Bethe-Salpeter equation in the matrix form. 
By introducing $4\times 4$ matrices 
for $\Gamma_{abcd}(\boldQ,\epsilon)$ and $\Pi_{abcd}(\boldQ,\epsilon)$, 
\begin{align}
&\Gamma=
\left(
\begin{array}{@{\,}cccc@{\,}}
\Gamma_{1111}(\boldQ,\epsilon) & \Gamma_{1112}(\boldQ,\epsilon) &
\Gamma_{1121}(\boldQ,\epsilon) & \Gamma_{1122}(\boldQ,\epsilon)\\
\Gamma_{1211}(\boldQ,\epsilon) & \Gamma_{1212}(\boldQ,\epsilon) & 
\Gamma_{1221}(\boldQ,\epsilon) & \Gamma_{1222}(\boldQ,\epsilon)\\
\Gamma_{2111}(\boldQ,\epsilon) & \Gamma_{2112}(\boldQ,\epsilon) & 
\Gamma_{2121}(\boldQ,\epsilon) & \Gamma_{2122}(\boldQ,\epsilon)\\
\Gamma_{2211}(\boldQ,\epsilon) & \Gamma_{2212}(\boldQ,\epsilon) & 
\Gamma_{2221}(\boldQ,\epsilon) & \Gamma_{2222}(\boldQ,\epsilon)
\end{array}
\right)\label{eq:Gamma-matrix},\\
&\Pi=
\left(
\begin{array}{@{\,}cccc@{\,}}
\Pi_{1111}(\boldQ,\epsilon) & \Pi_{1112}(\boldQ,\epsilon) &
\Pi_{1121}(\boldQ,\epsilon) & \Pi_{1122}(\boldQ,\epsilon)\\
\Pi_{1211}(\boldQ,\epsilon) & \Pi_{1212}(\boldQ,\epsilon) & 
\Pi_{1221}(\boldQ,\epsilon) & \Pi_{1222}(\boldQ,\epsilon)\\
\Pi_{2111}(\boldQ,\epsilon) & \Pi_{2112}(\boldQ,\epsilon) & 
\Pi_{2121}(\boldQ,\epsilon) & \Pi_{2122}(\boldQ,\epsilon)\\
\Pi_{2211}(\boldQ,\epsilon) & \Pi_{2212}(\boldQ,\epsilon) & 
\Pi_{2221}(\boldQ,\epsilon) & \Pi_{2222}(\boldQ,\epsilon)
\end{array}
\right),\label{eq:Pi-matrix}
\end{align}
we can express the Bethe-Salpeter equation Eq. (\ref{eq:BSeq}) as follows:
\begin{align}
\Gamma=\gamma_{\textrm{imp}}^{2}\Pi
+\gamma_{\textrm{imp}}\Pi\Gamma.\label{eq:BS-formal}
\end{align}
Solving this matrix equation, we obtain the formal solution,
\begin{align}
\Gamma=M^{-1}\gamma_{\textrm{imp}}^{2}\Pi,\label{eq:BS-formal-solution}
\end{align}
where $M^{-1}$ is the inverse matrix of $M_{abcd}$, given by
\begin{align}
M_{abcd}=\delta_{a,d}\delta_{b,c}-\gamma_{\textrm{imp}}\Pi_{abcd}(\boldQ,\epsilon).\label{eq:Mmatrix}
\end{align}
Second, we calculate $M^{-1}$. 
By using Eqs. (\ref{eq:Pi-matrix}) and (\ref{eq:Mmatrix}), 
we can express the $4\times 4$ matrix $M$ as follows:
\begin{align}
&M=
\left(
\begin{array}{@{\,}cccc@{\,}}
1-A & -C & -C & -F\\
-C & -F & 1-F & -D\\
-C & 1-F & -F & -D\\
-F & -D & -D & 1-B
\end{array}
\right),\label{eq:M-matrix}
\end{align}
where 
\begin{align}
A&=\gamma_{\textrm{imp}}\Pi_{1111}(\boldQ,\epsilon),\label{eq:A}\\
B&=\gamma_{\textrm{imp}}\Pi_{2222}(\boldQ,\epsilon),\label{eq:B}\\
C&=\gamma_{\textrm{imp}}\Pi_{1112}(\boldQ,\epsilon)=\gamma_{\textrm{imp}}\Pi_{1121}(\boldQ,\epsilon)
=\gamma_{\textrm{imp}}\Pi_{1211}(\boldQ,\epsilon)
=\gamma_{\textrm{imp}}\Pi_{2111}(\boldQ,\epsilon)\label{eq:C},\\
D&=\gamma_{\textrm{imp}}\Pi_{2221}(\boldQ,\epsilon)=\gamma_{\textrm{imp}}\Pi_{2212}(\boldQ,\epsilon)
=\gamma_{\textrm{imp}}\Pi_{2122}(\boldQ,\epsilon)
=\gamma_{\textrm{imp}}\Pi_{1222}(\boldQ,\epsilon)\label{eq:D},\\
F&=\gamma_{\textrm{imp}}\Pi_{1122}(\boldQ,\epsilon)=\gamma_{\textrm{imp}}\Pi_{1212}(\boldQ,\epsilon)
=\gamma_{\textrm{imp}}\Pi_{2112}(\boldQ,\epsilon)\notag\\
&=\gamma_{\textrm{imp}}\Pi_{1221}(\boldQ,\epsilon)=\gamma_{\textrm{imp}}\Pi_{2121}(\boldQ,\epsilon)
=\gamma_{\textrm{imp}}\Pi_{2211}(\boldQ,\epsilon).\label{eq:F}
\end{align}
To obtain $M^{-1}$, 
we need to calculate the cofactor matrix and determinant of $M$. 
After some algebra, 
we obtain
\begin{align}
M^{-1}
=\frac{1}{\textrm{det}M}
\left(
\begin{array}{@{\,}cccc@{\,}}
-1+2F+B & -C & -C & -F\\
-C & -F & -1+A+B+F & -D\\
-C & -1+A+B+F & -F & -D\\
-F & -D & -D & -1+A+2F
\end{array}
\right),\label{eq:M^-1-matrix}
\end{align}
where 
\begin{align}
\textrm{det}M=&-1+A+B+2F
=
\begin{cases}
-D_{\textrm{S}}(\epsilon)Q^{2}\tilde{\tau}(\epsilon)\ \ \ \ \ \ \ \ \ (\epsilon > 0),\\
-D_{\textrm{S}}(-\epsilon)Q^{2}\tilde{\tau}(-\epsilon)\ \  \ \ (\epsilon < 0).
\end{cases}\label{eq:detM}
\end{align}
In deriving Eq. (\ref{eq:detM})
we have used Eqs. (\ref{eq:A}), (\ref{eq:B}), (\ref{eq:F}) and Eq. (\ref{eq:Pi-approx}). 
Third, 
we combine Eqs. (\ref{eq:M^-1-matrix}), (\ref{eq:BS-formal-solution}) and (\ref{eq:Pi-matrix}). 
As a result, 
$\Gamma_{abcd}(\boldQ,\epsilon)$ is given by
\begin{align}
\Gamma_{abcd}(\boldQ,\epsilon)
=&-\frac{\gamma_{\textrm{imp}}}{\textrm{det}M}\gamma_{\textrm{imp}}\Pi_{abcd}(\boldQ,\epsilon)
=
\begin{cases}
\dfrac{u_{b\alpha}u_{c\alpha}u_{d\alpha}u_{a\alpha}\gamma_{\textrm{imp}}}
{D_{\textrm{S}}(\epsilon)Q^{2}\tilde{\tau}(\epsilon)}
\ (\epsilon > 0),\\[8pt]
\dfrac{u_{b\beta}u_{c\beta}u_{d\beta}u_{a\beta}\gamma_{\textrm{imp}}}
{D_{\textrm{S}}(-\epsilon)Q^{2}\tilde{\tau}(-\epsilon)}
\ \ (\epsilon < 0).
\end{cases}
\end{align}

\section{Derivation of Eq.(\ref{eq:anisotropy-final})}
In this appendix 
we derive Eq. (\ref{eq:anisotropy-final}) 
from Eq. (\ref{eq:anisotropy}) 
by calculating the dominant terms including the leading correction from $D/J$. 
Since the quantities on the right-hand side of Eq. (\ref{eq:anisotropy}) 
can be expressed in terms of $A(\boldq_{0})$, $B(\boldq_{0})$, 
$\partial A(\boldq_{0})/\partial \boldq_{0}$, and $\partial B(\boldq_{0})/\partial \boldq_{0}$, 
we first calculate the dominant terms of 
$\partial A(\boldq)/\partial \boldq$ and $\partial B(\boldq)/\partial \boldq$. 
From Eqs. (\ref{eq:Aq}) and (\ref{eq:Bq}) we obtain
\begin{align}
\frac{\partial A(\boldq)}{\partial q_{x}}
=&\frac{S}{2}(\sqrt{J^{2}+D^{2}}-J)\sin q_{x}
\sim \frac{S}{4}\frac{D^{2}}{J}\sin q_{x},\\
\frac{\partial A(\boldq)}{\partial q_{z}}
=&\ 0,\\
\frac{\partial B(\boldq)}{\partial q_{x}}
=&\frac{S}{2}(\sqrt{J^{2}+D^{2}}+J)\sin q_{x}
\sim SJ\sin q_{x}+\frac{S}{4}\frac{D^{2}}{J}\sin q_{x},\\
\frac{\partial B(\boldq)}{\partial q_{z}}
=&SJ\sin q_{z}.
\end{align}
Second, 
by using these equations, 
we estimate $e_{11}^{\alpha}(\boldq)$ and $e_{12}^{\alpha}(\boldq)$. 
The results are as follows:
\begin{align}
e_{11}^{x}(\boldq)
&\sim 2SJ\sin q_{x}B(\boldq)
\Bigl[1-\Bigl(\frac{D}{2J}\Bigr)^{2}\frac{A(\boldq)-B(\boldq)}{B(\boldq)}\Bigr],\\
e_{11}^{z}(\boldq)
&\sim 2SJ\sin q_{z}B(\boldq),\\
e_{12}^{x}(\boldq)
&\sim -2SJ\sin q_{x}A(\boldq)
\Bigl[1+\Bigl(\frac{D}{2J}\Bigr)^{2}\frac{A(\boldq)+B(\boldq)}{A(\boldq)}\Bigr],\\
e_{12}^{z}(\boldq)
&\sim -2SJ\sin q_{z}A(\boldq).
\end{align}
Third, by using these equations and Eq. (\ref{eq:tanh}), 
we rewrite the numerator and denominator in Eq. (\ref{eq:anisotropy}). 
We thus obtain  
\begin{align}
\frac{e_{11}^{x}(\boldq_{0})^{2}+e_{12}^{x}(\boldq_{0})^{2}\sinh^{2}2\theta_{\boldq_{0}}}
{e_{11}^{z}(\boldq_{0})^{2}+e_{12}^{z}(\boldq_{0})^{2}\sinh^{2}2\theta_{\boldq_{0}}}
=1+\frac{1}{2}\Bigl(\frac{D}{J}\Bigr)^{2}
\frac{2A(\boldq_{0})^{2}-B(\boldq_{0})^{2}+A(\boldq_{0})B(\boldq_{0})}
{2A(\boldq_{0})^{2}-B(\boldq_{0})^{2}}.\label{eq:anisotropy-pre}
\end{align}
Then 
the dominant terms of $A(\boldq_{0})$ and $B(\boldq_{0})$ are given by
$A(\boldq_{0})\sim 2SJ$ and 
$B(\boldq_{0})\sim -2SJ(1-\frac{q_{0}^{2}}{4})$; 
here we have approximated $\cos q_{0x}$ and $\cos q_{0z}$ 
as $\cos q_{0x}\sim 1-\frac{q_{0x}^{2}}{2}$ and $\cos q_{0z}\sim 1-\frac{q_{0z}^{2}}{2}$ 
and considered only the leading terms. 
By substituting these equations of $A(\boldq_{0})$ and $B(\boldq_{0})$
into Eq. (\ref{eq:anisotropy-pre}), 
we finally obtain Eq. (\ref{eq:anisotropy-final}). 

\end{widetext}



\begin{thebibliography}{99}
\bibitem{remark}
The time-reversal symmetry of the Dzyaloshinsky-Moriya interaction can been seen 
from the following argument. 
The Dzyaloshinsky-Moriya interaction is expressed as 
$\sum_{\boldi,\boldj}\boldD_{\boldi\boldj}\cdot 
(\hat{\boldS}_{\boldi}\times\hat{\boldS}_{\boldj})$. 
First, $(\hat{\boldS}_{\boldi}\times\hat{\boldS}_{\boldj})$ is symmetric 
about time reversal because 
the time-reversal operator~\cite{Sakurai} $\Theta$ satisfies 
$\Theta \hat{\boldS}_{\boldi}=-\hat{\boldS}_{\boldi}$, i.e.,  
$\Theta(\hat{\boldS}_{\boldi}\times\hat{\boldS}_{\boldj})
=(\hat{\boldS}_{\boldi}\times\hat{\boldS}_{\boldj})$. 
The coefficient $\boldD_{\boldi\boldj}$ also holds time-reversal symmetry.  
This can be seen, for example, from Eq. (2.8) of Ref. \onlinecite{DM-Moriya} 
by using the following three facts: 
the coefficient includes not only the imaginary unit $i$ 
but also the matrix element of the orbital angular momentum $\hat{\boldL}_{\boldi}$; 
$\Theta$ satisfies $\Theta i=-i$ and 
$\Theta\hat{\boldL}_{\boldi}=-\hat{\boldL}_{\boldi}$; 
there is no sign change in the other quantities of the coefficient under time-reversal operation.
\bibitem{remark2}
The relation between the band degeneracy and the time-reversal symmetry 
can be shown by analyzing the relation between the first and the second terms 
of the first line in Eq. (\ref{eq:H0-mag-band}) 
under time reversal. 
By applying the time-reversal operator~\cite{Sakurai} $\Theta$ 
to the first term, 
we obtain 
$\Theta[\epsilon(\boldq)\hat{\gamma}_{\boldq}^{\dagger}\hat{\gamma}_{\boldq}]
=\epsilon(-\boldq)\hat{\gamma}_{-\boldq}\hat{\gamma}_{-\boldq}^{\dagger}
=\epsilon(\boldq)\hat{\gamma}_{-\boldq}\hat{\gamma}_{-\boldq}^{\dagger}$. 
Thus time-reversal symmetry relates 
the magnons for $\boldq$ and $-\boldq$ in our spiral magnet.
\end{thebibliography}
\end{document}